\documentclass[12pt]{article}
\usepackage{amsmath,graphicx}
\usepackage{amssymb}
\usepackage{hyperref}

\font\mybb=msbm10 at 12pt
\def\bb#1{\hbox{\mybb#1}}

\def\C {\bb{C}}

\def\cA {{\cal A}}
\def\cF {{\cal F}}
\def\cO {{\cal O}}
\def\cN {{\cal N}}

\def\tr{\mathop{\rm tr}\nolimits}

\makeatletter
\@addtoreset{equation}{section}
\@addtoreset{footnote}{section}
\makeatother

\newcommand{\be}{\begin{equation}}
\newcommand{\ee}{\end{equation}}
\newcommand{\wt}{\widetilde}
\newcommand{\wh}{\widehat}
\newcommand{\ol}{\overline}

\newcommand{\ra}{\rightarrow}

\newcommand{\lra}{\leftrightarrow}
\newcommand{\nn}{\nonumber}
\newcommand{\half}{\frac{1}{2}}
\newcommand{\del}{\partial}

\newcommand{\bra}[1]{\left\langle\, #1\,\right|}

\newcommand{\f}{\frac}
\newcommand{\p}{\partial}
\newcommand{\m}{{\mu}}
\newcommand{\n}{{\nu}}
\newcommand{\G}{{\Gamma}}

\newcommand{\eb}{{\overline{\epsilon}}}

\def \bea {\begin{eqnarray}}
\def \eea {\end{eqnarray}}
\def \beal#1 {\begin{align}#1\end{align}}

\def\matt[#1,#2,#3,#4]{\left(%
\begin{array}{cc} #1 & #2 \\ #3 & #4 \end{array} \right)} 

\begin{document}

\begin{flushright}
\hfill{YITP-17-113}
\end{flushright}
\begin{center}
\vspace{2ex}
{\Large {\bf 
Supersymmetric Gauge Theory with Space-time\\
\vspace{1ex}
Dependent Couplings
}}

\vspace*{5mm}
{\sc Jaewang Choi}$^{a}$\footnote{e-mail:
 {\tt jchoi@yukawa.kyoto-u.ac.jp}},
{\sc Jos\'e J. Fern\'andez-Melgarejo}$^{a,b}$\footnote{e-mail:
 {\tt josejuan@yukawa.kyoto-u.ac.jp}}
~and~
{\sc Shigeki Sugimoto}$^{a,c}$\footnote{e-mail:
 {\tt sugimoto@yukawa.kyoto-u.ac.jp}}

\vspace*{4mm} 

\hspace{-0.5cm}
{\it {$^{a}$
Center for Gravitational Physics, Yukawa Institute for Theoretical
 Physics,\\ Kyoto University, Kyoto 606-8502, Japan
}}

\vspace*{1mm}

{\it {$^{b}$
Departamento de F\'isica, Universidad de Murcia,\\
Campus de Espinardo,  30100 Murcia, Spain
}}

\vspace*{1mm}

{\it {$^{c}$
Kavli Institute for the Physics and Mathematics of the Universe (WPI),\\
 The University of Tokyo, Kashiwanoha, Kashiwa 277-8583, Japan
}}\\ 

\end{center}

\vspace*{.3cm}
\begin{center}
{\bf Abstract}
\end{center}

We study deformations of ${\cal N}=4$ supersymmetric Yang-Mills theory
with couplings and masses depending on space-time. The conditions to
preserve part of the supersymmetry are derived and a lot of solutions
of these conditions are found.
The main example is the case with $ISO(1,1)\times SO(3)\times SO(3)$
symmetry, in which couplings, as well as masses and the theta parameter,
can depend on two spatial coordinates. In the case in which $ISO(1,1)$
is enhanced to $ISO(1,2)$, it reproduces the supersymmetric
Janus configuration found by Gaiotto and Witten. When
$SO(3)\times SO(3)$ is enhanced to $SO(6)$, it agrees with
the world-volume theory of D3-branes embedded in F-theory
(a background with 7-branes in type IIB string theory).
We have also found the general solution of the supersymmetry conditions
for the cases with $ISO(1,1)\times SO(2)\times SO(4)$ symmetry.
Cases with time dependent couplings and/or masses are also
considered.

\thispagestyle{empty}

\newpage

\tableofcontents

\section{Introduction}

Quantum field theories (QFT) usually contain constant parameters
such as gauge couplings, Yukawa couplings, theta parameters,
mass parameters, etc.
We are also familiar with QFT with some parameters
depending on space-time. QFT in curved space-time and QFT
with background fields or space-time dependent sources
of some operators are such examples. QFT with varying
couplings\footnote{Here, we regard masses, as well as
other parameters in the action, as couplings.
}
may be less familiar, probably because there is no clear observational
evidence suggesting that such parameters in the Standard Model are not
constant.

However, in string theory, they are obtained as
the values of background fields and, in this sense,
there is no conceptual difference between putting QFT
in a curved space-time and making the couplings space-time dependent.
In fact, there are a lot of examples. The Newton constant and
gauge couplings are related to the value of the dilaton field $\phi$.
In type IIB string theory, it is combined with
the Ramond-Ramond (RR) 0-form field $C_0$,
which corresponds to an analog of the theta parameter,
to have a complex coupling $\tau=ie^{-\phi}+C_0$.
One way to make it vary in space is to include D7-branes,
or more generally $[p,q]$ 7-branes. Regarding $\tau$ as the modulus
of a torus, we can think of uplifting the 10-dimensional space-time
with 7-branes to an elliptic fibered 12-dimensional space-time,
which is called F-theory \cite{Vafa:1996xn}.
Another interesting example of varying coupling is the so-called
Janus configuration, in which the coupling depends on one of the spatial
coordinates \cite{Bak:2003jk,Clark:2004sb}.

In this paper, we consider deformations of ${\cal N}=4$ supersymmetric
Yang-Mills (SYM) theory with varying couplings as typical
examples of QFT with space-time dependent parameters.
In particular, we investigate the conditions to preserve
part of the supersymmetry (SUSY). This is a natural extension
of the work to find curved space-times preserving SUSY,
for which systematic methods using, \emph{e.g.}, topological twist
\cite{Witten:1988ze}
or supergravity \cite{Festuccia:2011ws} have been developed.
We are of course not the first ones to consider this class of
theories. Supersymmetric Janus configurations were studied
in \cite{Clark:2005te,DHoker:2006vfr,DHoker:2006qeo,Gomis:2006cu,
DHoker:2007zhm,DHoker:2007hhe,Gaiotto:2008sd,Suh:2011xc}.
Their generalization to the configurations with couplings depending on more
than one direction was also investigated in \cite{Kim:2008dj,Kim:2009wv}.
SUSY configurations in ${\cal N}=4$ SYM with varying couplings have
been studied, for instance, in
\cite{Ganor:2010md,Ganor:2014pha,Martucci:2014ema,Maxfield:2016lok,
Lawrie:2016axq,Couzens:2017way,Assel:2016wcr}
(see also \cite{Harvey:2007ab,Buchbinder:2007ar,Harvey:2008zz} for earlier works).
Time dependent couplings have also been considered, for instance, in
\cite{Chu:2006pa,Das:2006dz,Lin:2006ie,Das:2006pw,Chu:2007um,
Awad:2007fj,Awad:2008jf}. 

Our approach here is perhaps the most rudimentary one. We write down all
the possible deformations in the action of ${\cal N}=4$ SYM and the SUSY
variation, and find the conditions to preserve part of the SUSY by
performing the SUSY variation. Although the calculation is a bit
tedious, it is straightforward and easy to understand the details.
We hope it will provide a useful guide for further analyses and
generalizations to other SUSY QFT.

Many of the examples considered in this paper
can be realized as the world-volume gauge theory
on probe D3-branes embedded in some non-trivial backgrounds
in type IIB string theory. This system is related by duality to
M5-branes wrapped on a torus with varying modulus in M-theory.
This gives a 6-dimensional description of the 4-dimensional QFT
with its complex coupling identified with the modulus of the
torus that corresponds to the extra two dimensions,
which is analogous to the idea of F-theory mentioned above.
The SUSY condition for the deformed ${\cal N}=4$ SYM
should be related to the conditions to preserve SUSY for
the D3-branes in the supergravity background.
In this way, our field theoretical analysis contains some
information on the supergravity background. As we will mention
in Section \ref{SO6}, it is possible to extract part of
the equations of motion, including the Einstein equation,
of supergravity for the case of an F-theory configuration.

The organization of the paper is as follows.
In Section \ref{S02}, after fixing our notations and ansatz for the action
and SUSY variation, the conditions to preserve SUSY are derived.
The details for the calculation are summarized in Appendix
\ref{appSUSYcond}.
It turns out that one of the SUSY conditions only has a trivial
solution if we impose certain symmetry.
Appendix \ref{app-eIJK} provides an explanation for this fact.
In Sections \ref{Ex33} and \ref{S04}, we demonstrate how our formalism works
by examining some explicit examples. We will analyze 
the case with $ISO(1,1)\times SO(3)\times SO(3)$ symmetry
in detail in Section \ref{Ex33}. We show in Section \ref{more}
that the SUSY conditions reduce to two simple equations (\ref{gh}) and
(\ref{hgk}).
As shown in Section \ref{sol1} and Section \ref{sol2},
large classes of solutions of these equations are found.
An F-theory configuration corresponding to the case with 
$ISO(1,1)\times SO(6)$ symmetry and
the supersymmetric Janus configuration found in \cite{Gaiotto:2008sd}
are obtained as special solutions.
The case with $ISO(1,1)\times SO(2)\times SO(4)$ symmetry
is studied in Section \ref{Ex24}, in which the general solution for this
case is found. Some examples with time dependence
are discussed in Section \ref{timedep}.
In Section \ref{conclusion}, we summarize the paper
and discuss some future directions.

\section{SUSY conditions for $\cN=4$ SYM with varying couplings}
\label{S02}

\subsection{Notations and Ansatz}
\label{NotationAnsatz}

In this paper, we consider $\cN=4$
$SU(N)$ supersymmetric Yang-Mills (SYM) theory
in a curved background with space-time dependent
gauge coupling $g_{\rm YM}(x^\mu)$ and theta parameter $\theta(x^\mu)$.
The leading order action is
\begin{eqnarray}
 S_0=\int d^4 x \sqrt{-g}\,\tr
\left\{\frac{1}{g_{\rm YM}^2}
\left(
-\frac{1}{2}g^{II'}g^{JJ'}F_{IJ}F_{I'J'}+i\ol\Psi\Gamma^I
D_I\Psi
\right)+\frac{\theta}{32\pi^2}
\epsilon^{\mu\nu\rho\sigma}F_{\mu\nu}F_{\rho\sigma}
\right\}
\ ,
\label{action0}
\end{eqnarray}
where $I,J=0,1,\cdots,9$ and
\begin{eqnarray}
F_{\mu\nu}&\equiv&\del_\mu A_\nu-\del_\nu A_\mu+i[A_\mu,A_\nu]\ ,
\nn\\
F_{\mu A}&=&-F_{A\mu}\equiv \del_\mu A_A+i[A_\mu,A_A]\equiv
D_\mu A_A\ ,
\\
F_{AB}&\equiv& i[A_A,A_B]\ ,
\nn
\end{eqnarray}
\begin{eqnarray}
D_\mu \Psi\equiv\del_\mu\Psi+i[A_\mu,\Psi]
+\frac{1}{4}\omega_\mu^{~\hat\nu\hat\rho}\Gamma_{\hat\nu\hat\rho}\Psi
\ ,~~~
D_A \Psi\equiv i[A_A,\Psi]
\ ,
\label{covder}
\end{eqnarray}
with $\mu,\nu=0,1,2,3$ and $A,B=4,\cdots,9$.
Here we are using the 10-dimensional notation,\footnote{
Our notation is similar to (but, not exactly the same as)
that used in \cite{Gaiotto:2008sd}.
}
in which
the gauge field $A_\mu$ and 6 adjoint scalar fields $A_A$
are combined into a 10-dimensional gauge field $A_I$, though
it depends only on the 4-dimensional space-time.
The 10-dimensional metric $g_{IJ}$ is assumed to be of the form
\begin{eqnarray}
 ds^2=g_{IJ}dx^Idx^J=g_{\mu\nu}(x^\rho)dx^\mu dx^\nu
+\delta_{AB}dx^Adx^B\ ,
\label{metric}
\end{eqnarray}
and $g^{IJ}$ is its inverse.\footnote{One may introduce
a non-trivial internal part of the metric $g_{AB}(x^\mu)$. However,
for our purpose, we can set it to be $g_{AB}(x^\mu)=\delta_{AB}$
without loss of generality. This is because
$g_{AB}$ in the action (\ref{action1}) can always be replaced with
$\delta_{AB}$ by redefinition of the scalar fields and the couplings
$a$, $d^{IJA}$, $m^{AB}$ and $M$. The off-diagonal components
 $g_{\mu A}$ are set to be zero for simplicity.
}
The indices are lowered or raised by this metric and its inverse.
$\epsilon^{\mu\nu\rho\sigma}$ is the 4-dimensional epsilon tensor
with $\epsilon^{0123}=1/\sqrt{-g}$, where
 $\sqrt{-g}\equiv\sqrt{-\det(g_{IJ})}=\sqrt{-\det(g_{\mu\nu})}$.

The fermion field $\Psi$ is written as a 10-dimensional
negative chirality Majorana-Weyl
spinor, which is equivalent to 4 Weyl spinor fields in
4-dimensional space-time.
It is a real 32 component spinor satisfying
\begin{eqnarray}
 \Gamma^{(10)}\Psi =-\Psi\ ,
\label{chirality}
\end{eqnarray}
where $\Gamma^{(10)}\equiv \Gamma^{\hat 0}\Gamma^{\hat 1}\cdots
\Gamma^{\hat 9}$
is the 10-dimensional chirality operator.\footnote{
This should not be confused with
$\Gamma^{10}=\Gamma^{1}\Gamma^{0}$.
}
Here, the gamma matrices $\Gamma^{\hat I}$ ($\hat I=0,1,\cdots,9$)
are 10-dimensional gamma matrices which are realized as $32\times 32$
real matrices satisfying
 $\{\Gamma^{\hat I},\Gamma^{\hat J}\}=2\eta^{\hat I\hat J}$,
where $(\eta^{\hat I\hat J})\equiv {\rm diag}(-1,+1,\cdots,+1)$
is the Minkowski metric 
(see Appendix \ref{Gamma-matrices} for our notations and useful formulas
for the gamma matrices).
The hatted indices $(\hat I,\hat J,\cdots)$
are those for the local Lorentz frame that
can be converted to the curved indices $(I,J,\cdots)$
by contracting them
with vielbeins $e_{\hat I}^I$
as $\Gamma^{I}=e_{\hat I}^I\Gamma^{\hat I}$.
The Dirac conjugate $\ol\Psi$ is defined
by $\ol\Psi\equiv \Psi^T\Gamma^{\hat 0}$.
Gamma matrices with more than one index are
anti-symmetrized products of the gamma matrices defined as
\begin{eqnarray}
 \Gamma^{I_1I_2\cdots I_n}=
 \Gamma^{[I_1}\Gamma^{I_2}\cdots\Gamma^{I_n]}=
\frac{1}{n!}
\sum_{\sigma\in S_n} {\rm sgn}(\sigma)
\Gamma^{I_{\sigma(1)}}\Gamma^{I_{\sigma(2)}}\cdots\Gamma^{I_{\sigma(n)}}
\ .
\label{antisymG}
\end{eqnarray}
$\omega_\mu^{~\hat\nu\hat\rho}$ in the covariant derivative $D_\mu$
in (\ref{covder}) is the spin connection:
\begin{eqnarray}
 \omega_{\mu\hat\nu\hat\rho}=\half e^{\nu'}_{\hat\nu}(\del_\mu e_{\nu'\hat\rho}
-(\del_{\nu'} g_{\mu\mu'})e^{\mu'}_{\hat\rho})-(\hat\nu\lra\hat\rho)\ .
\end{eqnarray}
In our notation,
$D_\mu$ denotes the covariant derivative including
gauge field $A_\mu$, spin connection
$\omega_\mu^{~\hat\nu\hat\rho}$ and Levi-Civita connection
$\Gamma_{\mu\nu}^{\rho}$,
depending on the field on which it acts.

When the metric is flat and the couplings ($g_{\rm YM}$ and $\theta$)
are constant, we know that the action \eqref{action0} is invariant under the
supersymmetry (SUSY)
transformation with 16 independent SUSY
parameters.\footnote{In this
paper we will not consider the special conformal supersymmetry.}
If the metric and/or the couplings are not constant,
SUSY is in general completely broken.
In order to maintain part of SUSY, we have to add
additional terms to the action (\ref{action0}) and the
SUSY transformation has to be modified accordingly.
The action that we consider is
\begin{multline}
 S=\int d^4 x \sqrt{-g}\,a\tr
\bigg\{
-\frac{1}{2}\,g^{II'}g^{JJ'}F_{IJ}F_{I'J'}+
i\,\ol\Psi\Gamma^I D_I\Psi
+\frac{c}{4}\,\epsilon^{\mu\nu\rho\sigma}F_{\mu\nu}F_{\rho\sigma}
\\
-d^{IJA}F_{IJ}A_A-\frac{m^{AB}}{2}A_AA_B
-i\ol\Psi M \Psi
\bigg\}
\ ,
\label{action1}
\end{multline}
where $a$, $c$, $d^{IJA}$, $m^{AB}$ are real parameters
and $M$ is a real $32\times 32$ matrix
that may depend on the space-time coordinates $x^\mu$.
$a$ and $c$ are related to $g_{\rm YM}$ and $\theta$
in (\ref{action0}) by
\begin{eqnarray}
 a=\frac{1}{g_{\rm YM}^2}\ ,~~~c=\frac{g_{\rm YM}^2\theta}{8\pi^2}\ .
\label{couplings}
\end{eqnarray}
We also use the complex coupling defined by
\begin{eqnarray}
 \tau\equiv \frac{\theta}{2\pi}+\frac{4\pi i}{g_{\rm YM}^2}
=4\pi a (c+i)\ .
\label{tau}
\end{eqnarray}

We impose the following conditions for the parameters
$d^{IJA}$ and $m^{AB}$,
which can be done without loss of generality:\footnote{
Square and round brackets on indices
indicate anti-symmetrization and symmetrization of the
indices, respectively. For example,
\begin{eqnarray}
d^{[ABC]}&\equiv& \frac{1}{3!}(d^{ABC}+d^{BCA}+d^{CAB}
-d^{BAC}-d^{CBA}-d^{ACB})\ ,
\nn\\
d^{(ABC)}&\equiv& \frac{1}{3!}(d^{ABC}+d^{BCA}+d^{CAB}
+d^{BAC}+d^{CBA}+d^{ACB})\ .
\nn
\end{eqnarray}
}
\begin{eqnarray}
 d^{IJ A}= -d^{JI A}\ ,~~~
 d^{\mu A B}= -d^{\mu BA}\ ,~~~
 d^{A B C}= d^{[ABC]}\ ,~~~
 m^{AB}=m^{BA}\ .
\label{d-and-m}
\end{eqnarray}
The second condition in (\ref{d-and-m}) can be imposed
because a term with
$d^{\mu (AB)}\tr(F_{\mu A}A_B)$
can be converted to a mass
term for $A_A$ via integration by parts.
We can further assume that the matrix
$(m^{AB})$ is diagonal, which can be realized, at least locally,
by using local $SO(6)_R$ transformation.

Note that $\ol\Psi M\Psi$ is non-vanishing only when
$\Gamma^{\hat 0}M$ is an anti-symmetric matrix that commutes with
$\Gamma^{(10)}$. The most general form of such a matrix is
\begin{eqnarray}
 M=m_{IJK}\Gamma^{IJK}\ ,
\label{M}
\end{eqnarray}
where $m^{IJK}$ is a real rank three
totally anti-symmetric tensor.

The action (\ref{action1}) is constructed by adding operators of
dimension\footnote{
Our analysis in this paper is classical and the anomalous dimensions
are not taken into account.
} less than 4 into the leading order action (\ref{action0}).
Although it is not the most general one,\footnote{
For example, an operator like $\tr(A_{(A}A_BA_{C)})$ is not included.
} it is general enough to cancel
the SUSY variations of the leading order action, as we will see in Section
\ref{SUSYcond} and Appendix \ref{appSUSYcond}.

The ansatz for the SUSY transformation is
\begin{eqnarray}
\delta_\epsilon A_I&=&i\ol\epsilon\,\Gamma_I\Psi\ ,
\nn\\
\delta_\epsilon \Psi&=&
 \frac{1}{2}(F_{IJ}\Gamma^{IJ}+A_AB^A)\,\epsilon\ ,
\\
\delta_\epsilon \ol\Psi&=&
\frac{1}{2}\ol\epsilon\,(-F_{IJ}\Gamma^{IJ}+A_A\ol B^A)\ ,
\nn
\label{SUSYtr}
\end{eqnarray}
where $B^A$ are real $32\times 32$ matrices acting on the spinor indices
that commute with the chirality operator $\Gamma^{(10)}$,
and $\ol B^A\equiv -\Gamma^{\hat 0}(B^A)^T\Gamma^{\hat 0}$.
The SUSY parameter $\epsilon$ is a 10-dimensional negative chirality
Majorana-Weyl spinor.
$B^A$ and $\epsilon$ may also depend on the space-time coordinates.

\subsection{SUSY conditions}
\label{SUSYcond}

The goal of this section is to determine the conditions under which the
action (\ref{action1}) is invariant with respect to the SUSY
transformations (\ref{SUSYtr}) for a non-zero $\epsilon$. The approach
that we follow in this work is straight. We firstly calculate the
variation of the deformed action \eqref{action1} with respect to the
deformed transformations \eqref{SUSYtr}. Then, by imposing the vanishing
of the variation, we obtain several constraints on the deformation
parameters and the SUSY parameter $\epsilon$.\footnote{
A similar analysis with a more elegant approach using supergravity \`a
la Festuccia-Seiberg \cite{Festuccia:2011ws} has been given in
\cite{Maxfield:2016lok}. 
However, a detailed comparison of the results still remains.
}
Here, we provide an outline of the derivation
and leave the details to Appendix \ref{appSUSYcond}.

Applying the SUSY variation (\ref{SUSYtr}) to
the action (\ref{action1}), we get
\begin{align}
\delta_\epsilon S
=&
\int d^4 x \sqrt{-g}\,a\,
\tr
\Bigg\{
(i\ol\epsilon\,\Gamma_I\Psi)
\Big[
-2g^{II'}g^{JJ'}
D_J F_{I'J'}
+\left(
-2a^{-1}D_\mu(a\,d^{I\mu A})
-m^{IA}\right) A_A
\nn\\
&\
\qquad\qquad\qquad\qquad\qquad\qquad
+\left(-2g^{IJ}g^{\mu K}
a^{-1}\del_\mu a
-a^{-1}\del_\nu(ac)\,\epsilon^{\nu IJK}
-3d^{[IJK]}\right)
F_{JK}
\Big]
\nn\\
&\
\qquad\qquad\qquad\quad
+
i\,\ol\epsilon\,(-F_{JK}\Gamma^{JK}+A_A\ol B^A)
\left(
\Gamma^I D_I\Psi
-\wt M \Psi
\right)
\Bigg\}
+(\mbox{total derivative terms})\, ,
\label{susyS0}
\end{align}
where
$\wt M\equiv M-\half\Gamma^\mu\del_\mu\log a$.
In this expression, $d^{JKI}$ and $m^{IA}$ can be non-zero only if
$I=4\sim 9$, and $\epsilon^{IJKL}$ can be non-zero only for
$I,J,K,L=0\sim 3$.

It can be shown after some calculation that (\ref{susyS0}) vanishes
(up to surface terms) if and only if the parameters satisfy the
following two conditions (see Appendix \ref{derivation1} for more
details):
\begin{align}
D_\mu\ol\epsilon\,\Gamma^{IJ}\Gamma^\mu
=&\
\ol\epsilon\left(
\ol{B}^{[J}\Gamma^{I]}
-\Gamma^{IJ\mu}a^{-1}\del_\mu a
+(a^{-1}\del_\nu(ac)\,\epsilon^{\nu IJK}
+3\,d^{[IJK]})\Gamma_{K}
-\Gamma^{IJ}\wt M
\right)\ ,
\label{SUSYcond1}
\\
D_\mu (\ol\epsilon\,\ol{B}^{A})\Gamma^\mu
=&\
\ol\epsilon
\left(-2a^{-1}\Gamma_ID_\mu (a\,d^{I\mu A})-m^{AB}\Gamma_B 
-\ol{B}^A\left(M+\half\Gamma^\mu\del_\mu\log a\right)
\right)
\ .
\label{SUSYcond2}
\end{align}
Conditions (\ref{SUSYcond1}) and (\ref{SUSYcond2}) correspond to
cancellation of terms with dimension
$\frac72$ and $\frac52$ operators, respectively. Following
\cite{Gaiotto:2008sd}, we call them first order and second order equations,
respectively.

Further algebra shows that the first order equation (\ref{SUSYcond1})
is equivalent to the following conditions (see Appendix
\ref{derivation2} for the derivation):
\begin{align}
0
=&\
\ol\epsilon\, e^{I'J'K'}
\Gamma_{K'}
\left(
P_{I'J'}^{~~~IJ}
-\delta_{I'}^I\delta_{J'}^J
\right)\ ,
\label{SUSYcond1-e}
\\
0 
=&\
\ol\epsilon\,\bigg(\frac{1}{72}e^{IJK}\Gamma_{IJK}
-
\frac{1}{2}\Gamma^{\mu}\del_\mu\log a
-\left(
\frac{1}{16}e_{\mu JK}-3m_{\mu JK}\right)\Gamma^{\mu JK}
-M
\bigg)\ ,
\label{am2}
\\
\ol\epsilon\,
\ol B^{A}
=&\ 
\ol\epsilon\left(F\Gamma^A+
\left(
-\frac{1}{4}e^{AJK}+12 m^{AJK}
\right)\Gamma_{JK}
\right)
\ ,
\label{Beq2}
\\
\del_\mu\ol\epsilon
=&\
\ol\epsilon\cA_\mu\ ,
\label{Dmueq2}
\end{align}
where $F$ is a real $32\times 32$ matrix acting on the spinor indices
(see (\ref{F}) or (\ref{F2})),
and
\begin{eqnarray}
 P_{I'J'}^{~~~IJ}
&\equiv&
\frac{1}{72}\Gamma_{I'J'}\Gamma^{IJ}
+\frac{1}{4}\Gamma_{[I'}\Gamma^{[I}\delta_{J']}^{J]}\ ,
\label{proj}
\\
e^{IJK}&\equiv& a^{-1}\del_\nu(ac)\,\epsilon^{\nu IJK}
+3\,d^{[IJK]}+24\,m^{IJK}\ ,
\label{eIJK}
\\
\cA_\mu&\equiv& -\frac{1}{4}\left(
F\Gamma_\mu+\left(
-\frac{1}{4}e_{\mu\hat J\hat K}
+12m_{\mu\hat J\hat K}-\omega_{\mu\hat J\hat K}
\right)\Gamma^{\hat J\hat K}
\right)
\ .
\label{cA}
\end{eqnarray}
$P_{I'J'}^{~~~IJ}$ is a projection, as it satisfies
\begin{eqnarray}
 P_{IJ}^{~~I'J'} P_{I'J'}^{~~~KL}= P_{IJ}^{~~KL}\ .
\end{eqnarray}
In addition, for an arbitrary $G^I$, we have 
\begin{eqnarray}
 G^{[I'}\Gamma^{J']}P_{I'J'}^{~~~IJ}
=G^{[I}\Gamma^{J]}\ ,
~~~
P_{I'J'}^{~~~IJ} G_{[I}\Gamma_{J]}
=G_{[I'}\Gamma_{J']}\ .
\end{eqnarray}

Note that $\ol B^A$, $F$ and $\cA_\mu$ always appear in the combination
$\eb\ol B^A$, $\eb F$ and $\eb\cA_\mu$, respectively,
and hence we only need to determine them up to additions
of matrices that vanish when $\eb$ is multiplied from the left.
In particular, the SUSY transformation (\ref{SUSYtr}) is
determined once the right hand side of (\ref{Beq2}) is fixed.

Equation (\ref{Dmueq2}) can be solved when the integrability condition
\begin{eqnarray}
 \ol\epsilon\cF_{\mu\nu}=0\ ,
\label{eF}
\end{eqnarray}
with $\cF_{\mu\nu}\equiv \del_\mu\cA_\nu-\del_\nu\cA_\mu
+[\cA_\mu,\cA_\nu]$
is satisfied. Then, (\ref{Dmueq2}) can be formally solved as
\begin{eqnarray}
 \ol\epsilon(x)=\ol\epsilon_0\,
{\rm P}\exp\left(\int_{x_0}^x dx^\mu\cA_\mu\right)\ ,
\label{epsilon-sol}
\end{eqnarray}
where $\ol\epsilon_0$ is a constant spinor, ``${\rm P}\exp$'' denotes
the path ordered exponential (the ordering is taken from left to right) 
and $x_0$  is a fixed position.
The integrability condition (\ref{eF}) guarantees that
 (\ref{epsilon-sol}) is well-defined in a neighborhood of $x_0$.

Equation (\ref{SUSYcond1-e}) has a trivial solution $e^{IJK}=0$.
In fact, in all the examples we consider in the following sections,
one can show that $e^{IJK}=0$ is the only solution of
(\ref{SUSYcond1-e}) that is compatible with the imposed symmetry.
When $e^{IJK}=0$, (\ref{am2}) is simplified as
\begin{eqnarray}
0 =
\ol\epsilon\,\bigg(
\frac{1}{2}\Gamma^{\mu}\del_\mu\log a
-3m_{\mu JK}\Gamma^{\mu JK}
+M
\bigg)\ .
\label{am3}
\end{eqnarray}
By definition, $e^{IJK}=0$ is equivalent to
\begin{eqnarray}
0&=& a^{-1}\del_\mu(ac)\,\epsilon^{\mu\nu\rho\sigma}
+24\,m^{\nu\rho\sigma}\ ,
\label{delac}
\\
0&=& d^{[IJA]}+8\,m^{IJA}
\ .
\end{eqnarray}
Using (\ref{d-and-m}), the latter is written as
\begin{eqnarray}
d^{\mu\nu A}= -24\,m^{\mu\nu A} 
\ ,~~~
d^{\mu AB}=-12\,m^{\mu AB} 
\ ,~~~
 d^{ABC}=-8\,m^{ABC}\ .
\label{md}
\end{eqnarray}
Equation (\ref{delac}) can have a solution if and only if
\begin{eqnarray}
 \del_\nu(\sqrt{-g}\,a\,m^{\nu\rho\sigma})=0
\label{delm}
\end{eqnarray}
is satisfied.

In summary, we have reduced the conditions for the SUSY invariance of
the action \eqref{action1} to the equations \eqref{SUSYcond1} and
\eqref{SUSYcond2}, where the former can be split into the equations
\eqref{SUSYcond1-e}-\eqref{Dmueq2}. In the following section, we are
going to elaborate on solutions that preserve different symmetries.

\section{Example: $ISO(1,1)\times SO(3)\times SO(3)$}
\label{Ex33}

In order to demonstrate how to solve the SUSY conditions obtained in the
previous section, we consider the cases with
 $ISO(1,1)\times SO(3)\times SO(3)$ symmetry. Here, $ISO(1,1)$ is the
Poincar\'e group acting on $x^{0,1}$,
and the first and second $SO(3)$ act as rotation of
$x^{4,5,6}$ and $x^{7,8,9}$ in the 10-dimensional notation,
respectively. 
Although our analysis is purely field theoretical, our motivation
is in string theory. Consider a D3-D5-D7 system in the following table:
\begin{eqnarray}
\begin{tabular}{c|ccccccccccccc}
&0&1&2&3&4&5&6&7&8&9\\
\hline
D3&o&o&o&o\\
D5&o&o&o&&o&o&o\\
D7&o&o&&&o&o&o&o&o&o
\end{tabular}
\label{D3D5D7}
\end{eqnarray}
This configuration preserves part of the supersymmetry
as well as the $ISO(1,1)\times SO(3)\times SO(3)$ symmetry.
If we regard the D5 and D7-branes as a supergravity background and
the D3-brane as a probe embedded in it, the low energy effective theory
on the D3-brane is expected to be a deformation of the $\cN=4$ SYM that
we have discussed. We may replace the D5-brane
and D7-branes with $(p,q)$ 5-brane and $[p,q]$ 7-branes, respectively,
to have more complicated configurations preserving SUSY.

The case with D3-branes in the $(p,q)$ 5-brane backgrounds corresponds
to the supersymmetric Janus configuration considered in
\cite{Gaiotto:2008sd}.
D3-branes in the $[p,q]$ 7-brane backgrounds can be generalized
to D3-branes in F-theory configurations, which were recently analyzed
in \cite{Martucci:2014ema,Lawrie:2016axq,Couzens:2017way} (see also
\cite{Harvey:2007ab,Buchbinder:2007ar,Harvey:2008zz}). 
These configurations appear as special cases in our example
as we discuss separately in Sections \ref{GW} and \ref{SO6}, respectively.

\subsection{Ansatz and SUSY conditions}
\label{33ansatz}

We decompose the coordinates in four sectors:
$\alpha,\beta=0,1$; $i,j=2,3$; $a,b,c=4,5,6$; $p,q,r=7,8,9$.
The metric and the couplings are assumed to
depend only on $x^i$ and preserve the
$ISO(1,1)\times SO(3)\times SO(3)$ symmetry acting on
$\{x^\alpha\}\times\{x^a\}\times\{x^p\}$.
The metric (\ref{metric}) is of the form
\begin{eqnarray}
ds^2=
e(x^i)\eta_{\alpha\beta}dx^\alpha dx^\beta
+g_{ij}(x^i)dx^i dx^j+\delta_{ab}dx^adx^b
+\delta_{pq}dx^pdx^q\ .
\end{eqnarray}
Using the general coordinate transformation and
Weyl transformation\footnote{See appendix \ref{Weyl}.}
with appropriate rescaling of the fields,
we can assume $e(x^i)=1$ and
$g_{ij}(x^i)=e^{\Phi(x^i)}\delta_{ij}$ without loss of generality.

The form of $M$ consistent with the symmetry is
\begin{eqnarray}
 M
&=& 6\left(m_{01i}\Gamma^{0 1 i}+m_{456}\Gamma^{456}+m_{789}\Gamma^{789}\right)
\nn\\
&\equiv&\alpha_i\Gamma^i\Gamma^{01}
+\beta\Gamma^{456}
+\gamma\Gamma^{789}
\ . 
\end{eqnarray}
The non-zero components of $d^{IJA}$ and  $m^{AB}$ are
\begin{eqnarray}
 d^{abc}=\frac{v}{3}\epsilon^{abc}\ ,~~
 d^{pqr}=\frac{w}{3}\epsilon^{pqr}\ ,~~
 m^{ab}=r \delta^{ab}\ ,~~
 m^{pq}=\wt r \delta^{pq}\ .
\end{eqnarray}
We also define
\begin{eqnarray}
 q_i\equiv\del_i\log a\ .
\end{eqnarray}

The non-trivial components of $e^{IJK}$ defined in (\ref{eIJK}) are
\begin{eqnarray}
e^{01i}=a^{-1}\del_j(ac)\,\epsilon^{ji}
-4\alpha^i
\ ,~~~
e^{abc}=(v+4\beta)\epsilon^{abc}\ ,~~~
e^{pqr}=(w+4\gamma)\epsilon^{abc}\ .
\end{eqnarray}
In this case, one can show that the condition (\ref{SUSYcond1-e})
implies $e^{IJK}=0$ (see Appendix \ref{e-mixed})
and hence
\begin{eqnarray}
a^{-1}\del_j(ac)\,\epsilon^{ji}=4\alpha^i
\ ,~~~
v=-4\beta\ ,~~~
w=-4\gamma\ .
\label{albega}
\end{eqnarray}
The integrability condition (\ref{delm}) for the first equation
of (\ref{albega}) is
\begin{eqnarray}
g^{kj} \del_k(a\alpha_j)=0\ .
\label{delaalpha}
\end{eqnarray}

The non-zero components of the spin connection are
\begin{eqnarray}
 \omega_i^{~\hat 2\hat 3}= -\omega_i^{~\hat 3\hat 2}
=\half\epsilon_i^{~j}\del_j\Phi\ ,
\label{spinconn}
\end{eqnarray}
and $\cA_\mu$ defined in (\ref{cA}) is
\begin{eqnarray}
&&\cA_0= -\frac{1}{4}\left(
F\Gamma_0+
4\alpha_i\Gamma^{1i}
\right)
\ ,
~~~
\cA_1= -\frac{1}{4}\left(
F\Gamma_1-
4\alpha_i\Gamma^{0i}
\right)
\ ,
\nn\\
&&\cA_i= -\frac{1}{4}\left(
F\Gamma_i+
4\alpha_i\Gamma^{01}
-\epsilon_i^{~j}\del_j\Phi\Gamma^{\hat 2\hat 3}
\right)
\ .
\end{eqnarray}
The condition (\ref{Dmueq2}) with $\mu=0,1$ implies
\begin{eqnarray}
 0=\ol\epsilon (F+4\alpha_i\Gamma^{01i})\ .
\label{Falpha}
\end{eqnarray}
Using this the equation, $\cA_i$ in (\ref{Dmueq2}) with $\mu=i=2,3$
can be replaced with
\begin{eqnarray}
 \cA_i=\epsilon_i^{~j}\left(
-\alpha_j\Gamma^{01}+\frac{1}{4}\del_j\Phi
\right)\Gamma^{\hat 2\hat 3}\ ,
\end{eqnarray}
and (\ref{am3}), (\ref{Beq2}) and (\ref{Dmueq2}) become
\begin{eqnarray}
0
&=&
\ol\epsilon\left(
\Gamma^iq_i
-4\alpha_i\Gamma^{01i}
+2\beta\Gamma^{456}
+2\gamma\Gamma^{789}
\right)
\ ,
\label{1st1}
\\
\ol\epsilon\ol B^a
&=&\ol\epsilon\left(
-4\alpha_i\Gamma^{01ia}
+2\beta\epsilon^{abc}\Gamma_{bc}
\right)\ ,
\label{1st2}
\\
\ol\epsilon\ol B^p
&=&\ol\epsilon\left(
-4\alpha_i\Gamma^{01ip}
+2\gamma\epsilon^{pqr}\Gamma_{qr}
\right)\ ,
\label{1st3}
\\
\del_i\ol\epsilon&=&\ol\epsilon
\,\epsilon_i^{~j}
\left(-\alpha_j\Gamma^{01}
+\frac{1}{4}\del_j\Phi\right)\Gamma^{\hat 2\hat 3}
\ .
\label{1st4}
\end{eqnarray}
The integrability condition (\ref{eF}) for (\ref{1st4}) is
\begin{eqnarray}
0=\ol\epsilon\, g^{kj}\left(
4\del_k\alpha_j\Gamma^{01}-\del_k\del_j\Phi
\right)\ .
\label{alphaPhi}
\end{eqnarray}

The second order equation (\ref{SUSYcond2}) for this case is
\begin{eqnarray}
D_i(\ol\epsilon\ol{B}^{a})\Gamma^i
&=&
\ol\epsilon
\left(-r\Gamma^a -\ol{B}^a
\left(M+\half\Gamma^i q_i\right)
\right)
\ ,
\label{2nd1}
\\
D_i(\ol\epsilon\ol{B}^{p})\Gamma^i
&=&
\ol\epsilon
\left(-\wt r\Gamma^p -\ol{B}^p
\left(M+\half\Gamma^i q_i\right)
\right)
\ .
\label{2nd2}
\end{eqnarray}
Using (\ref{1st2})--(\ref{1st4}), one can show that (\ref{2nd1}) and
(\ref{2nd2}) are equivalent to
\begin{eqnarray}
0&=&r+\wt r
+g^{ij}q_iq_j+2g^{ij}\del_iq_j
-4(\beta^2+\gamma^2)
\ ,
\label{2nd3}
\\
0 &=&
\ol\epsilon
\bigg(r-\wt r
+4\del_i\beta\Gamma^{i456}
-4\del_i\gamma\Gamma^{i789}
+8\alpha_i\Gamma^{01i}
\left(
\beta\Gamma^{456}
-\gamma\Gamma^{789}\right)
\bigg)\ .
\label{2nd5}
\end{eqnarray}

Therefore, equations that we have to solve are
(\ref{delaalpha}), (\ref{1st1}), (\ref{alphaPhi})
and (\ref{2nd5}). Once we find $a$,
$\alpha_j$, $\beta$, $\gamma$, $\Phi$ and $r-\wt r$
satisfying these equations,
the other parameters can be easily obtained by
(\ref{albega}), (\ref{1st2}), (\ref{1st3})
and (\ref{2nd3}). It can be easily checked that these
SUSY conditions reduce to those given in \cite{Gaiotto:2008sd}
when the symmetry is enhanced to $ISO(1,2)\times SO(3)\times SO(3)$.

\subsection{Solutions of the SUSY conditions}

In this section we are going to elaborate on a prescription for the SUSY
parameter $\epsilon$ that simplifies the study of solutions. In
addition, we give two examples of solutions, where the latter is a
generalization of the former. 

\subsubsection{More on SUSY conditions}
\label{more}

Let us first try to solve (\ref{alphaPhi}).
Decomposing $\ol\epsilon$ as
\begin{eqnarray}
 \ol\epsilon= \ol\epsilon_++\ol\epsilon_-\ ,
\end{eqnarray}
with $\ol\epsilon_\pm \Gamma^{01}=\pm\ol\epsilon_\pm$,
(\ref{alphaPhi}) can be written as
\begin{eqnarray}
0=
\ol\epsilon_+ g^{kj}\left(
4\del_k\alpha_j-\del_k\del_j\Phi
\right)
-\ol\epsilon_- g^{kj}\left(
4\del_k\alpha_j+\del_k\del_j\Phi
\right)\ .
\end{eqnarray}
If $g^{kj}\del_k\del_j\Phi=0$ and $g^{kj}\del_k\alpha_j=0$,
both $\ol\epsilon_+$ and $\ol\epsilon_-$ can be non-zero.
However, if $g^{kj}\del_k\del_j\Phi\ne 0$, 
it has a non-trivial solution only if
\begin{eqnarray}
 g^{kj}\del_k(\del_j\Phi\mp 4\alpha_j)=0\ ,~~\ol\epsilon_\mp=0
\label{alphaPhi2}
\end{eqnarray}
are satisfied.
Therefore, when $\Phi$ is not a harmonic
function, the unbroken SUSY is inevitably chiral in two dimensions.
In the following, we impose (\ref{alphaPhi2}), though we do not
assume $g^{kj}\del_k\del_j\Phi\ne 0$. The general solution
for the case with $g^{kj}\del_k\del_j\Phi=0$ can be easily obtained by
taking a linear combination of a solution with
$\ol\epsilon=\ol\epsilon_+$ and that with $\ol\epsilon=\ol\epsilon_-$.

Equation (\ref{alphaPhi2}) can be solved (at least locally) if and only
if there exists a function $\varphi_\pm$ satisfying
\begin{eqnarray}
\pm\alpha_j-\frac{1}{4}\del_j\Phi=\epsilon_j^{~i} \del_i\varphi_\pm\ .
\label{alphaPhivarphi}
\end{eqnarray}
Then, the solution of (\ref{1st4}) is
\begin{eqnarray}
\ol\epsilon
=\ol\epsilon^0_\pm e^{\varphi_\pm \Gamma^{\hat 2\hat 3}}
=\ol\epsilon^0_\pm
\left(\cos\varphi_\pm+\Gamma^{\hat 2\hat 3}\sin\varphi_\pm\right)\ ,
\label{inteps0}
\end{eqnarray}
where $\ol\epsilon^0_\pm$ is a constant spinor satisfying
\begin{eqnarray}
\ol\epsilon^0_\pm\,\Gamma^{01} =\pm\ol\epsilon^0_\pm\ ,~~~
\ol\epsilon^0_\pm\,\Gamma^{(10)} =\ol\epsilon^0_\pm\ .
\label{eps0chiral}
\end{eqnarray}
The second condition in (\ref{eps0chiral}) follows from the chirality condition (\ref{chirality}).

This $\ol\epsilon_\pm^0$ belongs to the 8-dimensional Majorana-Weyl
representation of the $SO(8)$ subgroup of the 10-dimensional Lorentz group.
Since the operators acting on $\ol\epsilon$ in (\ref{1st1})
and (\ref{2nd5}) commute with $SO(3)\times SO(3)$
generators $\Gamma^{ab}$ and $\Gamma^{pq}$,
it is convenient to decompose it to two $(2,2)$ representations
of the $SO(3)\times SO(3)$ as
\begin{eqnarray}
 \ol\epsilon_\pm^0=\sum_{v=1}^4(f_v^0\bra{0;v} +f_v^1\bra{1;v})\ ,
\end{eqnarray}
where $v=1,2,3,4$ is an index labeling the 4-dimensional representation
($(2,2)$ representation) of $SO(3)\times SO(3)$,
and $f_v^0$ and $f_v^1$ ($v=1,2,3,4$) are real parameters.
The set $\{\bra{0;v},\bra{1;v}\,|\, v=1,2,3,4\}$ is the basis of the spinors
satisfying (\ref{eps0chiral}). They can be constructed
explicitly as follows. Let us define
\begin{eqnarray}
C_1\equiv\frac{1}{2}(\Gamma^{\hat2}+\Gamma^{456})\ ,~~~
C_2\equiv\frac{1}{2}(\Gamma^{\hat3}\mp\Gamma^{789})\ ,
\end{eqnarray}
which satisfy
\begin{eqnarray}
 \{C_s,C_t^\dag\}=\delta_{st}\ ,~~~ \{C_s,C_t\}=0\ , ~~~(s,t=1,2)\ .
\end{eqnarray}
Let $\bra{0;v}$ be a spinor satisfying
\begin{eqnarray}
\bra{0;v} C_s^\dag=0\ ,~~~(s=1,2)
\label{vac}
\end{eqnarray}
and $\bra{1;v}$ is defined as
\begin{eqnarray}
 \bra{1;v}\equiv\bra{0;v}C_1C_2\ .
\end{eqnarray}
Since all the operators acting on $\ol\epsilon_\pm^0$ in  (\ref{1st1})
and (\ref{2nd5}) do not mix the spaces with different index $v$, we can
assume 
\begin{eqnarray}
\ol\epsilon_\pm^0=f^0_v\bra{0;v}+f^1_v\bra{1;v}
\label{ep0}
\end{eqnarray}
with fixed $v$.
(The general solution is just a linear combination
of this type.)
Note that $\Gamma^{\hat2\hat3}=(C_1+C_1^\dag)(C_2+C_2^\dag)$ and 
\begin{eqnarray}
 \bra{0;v} \Gamma^{\hat 2\hat 3}= \bra{1;v}\ ,~~~
 \bra{1;v} \Gamma^{\hat 2\hat 3}= -\bra{0;v}\ .
\end{eqnarray}
Then, (\ref{ep0}) can be written as
\begin{eqnarray}
 \ol\epsilon^0_\pm=\eta \bra{0;v}e^{\xi\,\Gamma^{\hat2\hat3}}\ ,
\label{ep0-2}
\end{eqnarray}
where $\eta=\sqrt{(f_v^0)^2+(f_v^1)^2}$ and
$\xi=\arctan (f^1_v/f^0_v)$.
As a consequence, the number of remaining SUSY
is 4 in general, corresponding to the choice of $v=1,2,3,4$.
This agrees with what we expect from the brane configuration
(\ref{D3D5D7}).
If the phase $\xi$ can be chosen freely without changing the
action, the number of SUSY is enhanced to 8.
As mentioned above, if $\Phi$ satisfies the Laplace equation
$g^{kj}\del_k\del_j\Phi=0$ and
both $\ol\epsilon_+$ and $\ol\epsilon_-$
are allowed, the number of SUSY is doubled.
We will see some examples with the SUSY enhancement later.

Since $\xi$ can be absorbed by the shift of $\varphi_\pm$ in
(\ref{inteps0}), we set $\xi=0$ in the following.
Then, using (\ref{alphaPhivarphi}), (\ref{inteps0}) and
(\ref{ep0-2}), one can show that the SUSY condition (\ref{1st1}) is equivalent to
\begin{eqnarray}
\beta&=& e^{-\Phi}a^{-1/2}\left(
 \del_2\left(e^{\half\Phi}a^{-1/2}
\,\cos(2\varphi_\pm)
\right)
+\del_3\left(e^{\half\Phi}a^{-1/2}
\,\sin(2\varphi_\pm)
\right)\right)
\ ,
\\
\gamma&=&\mp e^{-\Phi}a^{1/2}\left(
-\del_2\left(e^{\half\Phi}a^{-1/2}
\,\sin(2\varphi_\pm)
\right)
+\del_3\left(e^{\half\Phi}a^{-1/2}
\,\cos(2\varphi_\pm)
\right)
\right)
\ ,
\end{eqnarray}
and (\ref{2nd5}) is equivalent to
\begin{align} 
\del_2\left(e^{\half\Phi}(
-\beta\cos(2\varphi_\pm)
\pm\gamma\sin(2\varphi_\pm)
)
\right)
+\del_3\left(e^{\half\Phi}(
\mp\gamma\cos(2\varphi_\pm)
-\beta\sin(2\varphi_\pm)
)
\right)
=&\
-\frac{1}{4}(r-\wt r)e^{\Phi}
\ ,
\\
\del_2\left(e^{\half\Phi}(
\pm\gamma\cos(2\varphi_\pm)
+\beta\sin(2\varphi_\pm)
)
\right)
+\del_3\left(e^{\half\Phi}(
-\beta\cos(2\varphi_\pm)
\pm\gamma\sin(2\varphi_\pm)
)
\right)
=&\
0
\ .
\end{align}

It is convenient to write these equations using
a complex coordinate $z\equiv\frac{1}{\sqrt{2}}(x^2+ix^3)$:
\begin{eqnarray}
\beta\pm i\gamma
&=&
\sqrt{2}\, e^{-\Phi}a^{1/2}
\del_z\left(e^{\half\Phi+i2\varphi_\pm}a^{-1/2}
\right)\ ,
\label{beta+igamma}
\end{eqnarray}
\begin{eqnarray}
\frac{1}{4}(r-\wt r)e^{\Phi}
&=&
\sqrt{2}\,\del_z
\left(e^{\half\Phi+i2\varphi_\pm}
(\beta\pm i\gamma)
\right)
\nn\\
&=&
2\del_z
\left(e^{-\half\Phi+i2\varphi_\pm}a^{1/2}
\del_z\left(e^{\half\Phi+i2\varphi_\pm}a^{-1/2}
\right)
\right)\ ,
\label{r-rt}
\end{eqnarray}
where we have used (\ref{beta+igamma}) in the last step.
$\beta$, $\gamma$ and $(r-\wt r)$ are determined by
(\ref{beta+igamma}) and the real part of (\ref{r-rt}).
The imaginary part of (\ref{r-rt}) gives
a non-trivial constraint:
\begin{eqnarray}
{\rm Im}\left[\del_z
\left(e^{-\half\Phi+i2\varphi_\pm}a^{1/2}
\del_z\left(e^{\half\Phi+i2\varphi_\pm}a^{-1/2}
\right)\right)\right]=0\ .
\end{eqnarray}
This equation can be solved if there exists a real function
$f(z,\ol z)$ satisfying
\begin{eqnarray}
\del_{\ol z}f
&=&
e^{-\half\Phi+i2\varphi_\pm}a^{1/2}
\del_z\left(e^{\half\Phi+i2\varphi_\pm}a^{-1/2}
\right)
\nn\\
&=&
e^{i4\varphi_\pm}\left(i2\del_z\varphi_\pm+
\half\del_z\left(\Phi-\log a
\right)\right)\ .
\label{feq}
\end{eqnarray}
This is equivalent to
\begin{eqnarray}
e^{-i2\varphi_\pm+\half(\Phi-\log a)}\del_{\ol z}e^f
=
\del_z\left(e^{i2\varphi_\pm+\half(\Phi-\log a)}\right)e^f\ .
\label{feq2}
\end{eqnarray}
The complex conjugate of this equation is
\begin{eqnarray}
\del_{\ol z}\left(e^{-i2\varphi_\pm+\half(\Phi-\log a)}\right)e^f
=
e^{i2\varphi_\pm+\half(\Phi-\log a)}\del_{z}e^f\ .
\end{eqnarray}
The sum of these two equations gives
\begin{eqnarray}
\del_{\ol z}\left(e^{-i2\varphi_\pm+\half(\Phi-\log a)}
e^f\right)
=
\del_{z}\left(e^{i2\varphi_\pm+\half(\Phi-\log a)}e^f
\right)\ ,
\end{eqnarray}
which is equivalent to
\begin{eqnarray}
{\rm Im}\left[
\del_{z}\left(e^{i2\varphi_\pm+\half(\Phi-\log a)+f}
\right)\right]=0\ .
\end{eqnarray}
This equation can be solved if there exists a real function $g(z,\ol z)$
satisfying
\begin{eqnarray}
e^{i2\varphi_\pm+\half(\Phi-\log a)}
=e^{-f}\del_{\ol z}g\ .
\label{fdelg}
\end{eqnarray}
Inserting this into (\ref{feq2}), we obtain
\begin{eqnarray}
\del_z g  \del_{\ol z}f+\del_z f\del_{\ol z}g=\del_z\del_{\ol z}g\ .
\label{gf}
\end{eqnarray}
If we are able to find real functions $g$ and $f$ satisfying this
relation, $\varphi_\pm$ and $(\Phi-\log a)$ are obtained by
(\ref{fdelg}). Equations (\ref{beta+igamma}) and (\ref{r-rt}) are
\begin{eqnarray}
\beta\pm i\gamma&=&\sqrt{2}e^{-\Phi}a^{1/2}\del_z(e^{-f}\del_{\ol z}g)
=\sqrt{2}e^{-\Phi}a^{1/2}e^{-f}\del_z g\del_{\ol z}f
\ ,
\label{beta+igamma2}
\\ 
r-\wt r&=&8e^{-\Phi}\del_z\del_{\ol z}f\ .
\label{r-wtr}
\end{eqnarray}

Note that (\ref{gf}) can also be written as
\begin{eqnarray}
{\rm Re}[ \del_z(e^{-2f}\del_{\ol z}g)]=0\ .
\end{eqnarray}
This equation can be solved if there exists a real function $h(z,\ol z)$
satisfying
\begin{eqnarray}
 e^{-2f}\del_z g=i\del_z h\ ,
\label{expfdg}
\end{eqnarray}
which implies
\begin{eqnarray}
 \del_z g\del_{\ol z} h+ \del_{\ol z} g\del_z h=0\ .
\label{gh}
\end{eqnarray}
This shows that the gradients of $g$ and $h$ are orthogonal to
each other. Conversely, if we are able to find real functions $g$ and
$h$ satisfying (\ref{gh}), $f$ is obtained as
\begin{eqnarray}
 e^{-2f}=i\frac{\del_z h}{\del_z g}
=-i\frac{\del_{\ol z} h}{\del_{\ol z} g}\ .
\label{fgh}
\end{eqnarray}
Equation (\ref{fdelg}) can also be written as
\begin{eqnarray}
e^{i4\varphi_\pm+\Phi-\log a}
=-i\del_{\ol z} h\del_{\ol z}g\ .
\label{dhdg}
\end{eqnarray}

In addition to these equations, we should also solve (\ref{delaalpha}).
Using (\ref{alphaPhivarphi}),  (\ref{delaalpha}) can be written as
\begin{eqnarray}
g^{ij}\del_j (a\del_i\Phi)+4\epsilon^{jk}\del_ja
\del_k\varphi_\pm
=0\ ,
\end{eqnarray}
which is equivalent to
\begin{eqnarray}
{\rm Re}\left[\del_{\ol z}( a\,\del_z(\Phi+4i\varphi_\pm)
)\right]
=0\ .
\end{eqnarray}
This equation can be solved if there exists a real function $k(z,\ol z)$
satisfying
\begin{eqnarray}
 a\,\del_z(\Phi-\log a+4i\varphi_\pm)+\del_z a
= a\,\del_z(\Phi+4i\varphi_\pm)
=i\del_z k\ .
\label{keq}
\end{eqnarray}
{}From the first equation of (\ref{albega}) and (\ref{alphaPhivarphi}),
we see that this $k$ is proportional to the theta parameter as
\begin{eqnarray}
 k=\mp ac=\mp\frac{\theta}{8\pi^2}\ ,
\label{k-ac}
\end{eqnarray}
up to an additive constant.

Using (\ref{fdelg}), (\ref{keq}) becomes
\begin{eqnarray}
2 a\del_z(-f+\log\del_{\ol z}g)+\del_z a
=i\del_z k\ ,
\label{afgk}
\end{eqnarray}
which can also be written as
\begin{eqnarray}
\del_z(a\del_{\ol z}h\del_{\ol z}g)
=i\del_{\ol z}h\del_{\ol z}g\del_zk\ .
\label{hgk}
\end{eqnarray}

In summary, the SUSY conditions are now reduced to a problem of
finding real functions $h$, $g$, $a$ and $k$ satisfying
 (\ref{gh}) and (\ref{hgk}). Then, the real function $f$
is obtained by (\ref{fgh}) and other parameters are determined
by (\ref{fdelg}) (or (\ref{dhdg})),
(\ref{beta+igamma2}), (\ref{r-wtr}) and (\ref{k-ac}).
Although we have not been able to find the general solution
of the SUSY conditions (\ref{gh}) and (\ref{hgk}),
a lot of non-trivial solutions have been found.
In the following subsections, we show some of the
explicit solutions.

\subsubsection{Solution 1}
\label{sol1}

First, we introduce new coordinates $(y^1,y^2)$ defined as
\begin{eqnarray}
 y^1\equiv \ol l(\ol z)+l(z)\ ,~~~
 y^2\equiv i (\ol l(\ol z)-l(z))\ ,
\end{eqnarray}
where $l(z)\in \C$ is a holomorphic function of
$z=\frac{1}{\sqrt{2}}(x^2+ix^3)$ and $\ol l(\ol z)$ is its complex
conjugate.
They are related to the original coordinates $(x^2,x^3)$
by a conformal transformation on the 2-dimensional plane.
Note that our ansatz explained in Section \ref{33ansatz}
is compatible with the conformal transformation\footnote{
Because we set $g_{ij}(x^i)=e^{\Phi(x^i)}\delta_{ij}$,
the general coordinate transformation of $(x^2,x^3)$
is not compatible with our ansatz. The conformal transformation
on the $(x^2,x^3)$-plane keeps this form invariant.
}
and hence
our results in the previous subsection are valid in the
coordinates $(y^1,y^2)$ as well. In fact, it is easy to see
that the equations (\ref{gh}) and (\ref{hgk}) are invariant under
the conformal transformation. Once one finds a solution,
one can generate new solutions by the conformal transformations.
In order to emphasize this point, we write down the solutions
that work for any choice of the holomorphic function $l(z)$,
rather than using this degree of freedom to simplify the
equations.

Since the condition (\ref{gh}) is equivalent to the statement
that the gradient of $g$ and $h$ are orthogonal to each other,
it is clear that it can be solved when $g$ and $h$ are of the form:
\begin{eqnarray}
g(z,\ol z)=G_1(y^1)\ ,~~~
h(z,\ol z)=H_2(y^2)\ ,
\end{eqnarray}
where $G_1$ and $H_2$ are real functions.
The subscripts of these functions
suggest which of the coordinates ($y^1$ or $y^2$)
they depend on.
Inserting these into (\ref{hgk}), we obtain
\begin{eqnarray}
 \del_z(a H_2'G_1')=i H_2'G_1' \del_z k\ ,
\label{aHG}
\end{eqnarray}
where the prime denotes the derivative, \emph{e.g.},
$G'_1\equiv\del_{y^1}G_1(y^1)$.

One can check that the following ansatz gives
a solution of (\ref{aHG}):
\begin{eqnarray}
a(z,\ol z)=\frac{L(y^1,y^2)}{G_1'(y^1)H_2'(y^2)}\ ,~~~
k(z,\ol z)=K_1(y^1)+K_2(y^2)\ ,
\label{ak}
\end{eqnarray}
where $K_1$ and $K_2$ are real functions satisfying
\begin{eqnarray}
 G_1'K_1'=\kappa_1\ ,~~~ H_2'K_2'=\kappa_2\ ,
\label{AKK}
\end{eqnarray}
with real constants $\kappa_i$ ($i=1,2$) and $A$ is
a real function defined as
\begin{eqnarray}
L(y^1,y^2)\equiv \kappa_2 G_1(y^1)-\kappa_1 H_2(y^2)\ .
\end{eqnarray}
Then, using
(\ref{2nd3}), (\ref{beta+igamma2}), (\ref{r-wtr}), (\ref{dhdg})
and (\ref{k-ac}),
we obtain the rest of the parameters:
\begin{align}
c
=&\
\mp\frac{G_1'H_2'}{L}(K_1+K_2)\ ,
\label{c-sol1}
\\
\Phi
=&\
\log\left(L\del_zl\del_{\ol z}\ol l\right)\ ,
\label{Philog}
\\
\varphi_\pm
=&\
\half{\rm Im}[\log\del_{\ol z}\ol l]\ ,
\\
\beta\pm i\gamma
=&\
\frac{1}{\sqrt{2L}}\left((\log G_1')'-i(\log H_2')'\right)\ ,
\\
r-\wt r
=&\
\frac{4}{L}\left((\log G_1')''-(\log H_2')''\right)\ ,
\\
r+\wt r
=&\
\frac{4}{L}\left((\log G_1')''+(\log H_2')''\right)
+\frac{2}{L^3}(\kappa_2^2 G_1'^2+\kappa_1^2 H_2'^2)
\ .
\label{rwtr-sol1}
\end{align}

\subsubsection{Solution 2}
\label{sol2}

Let us generalize the solutions in the previous subsection
by considering the following ansatz:
\begin{eqnarray}
g(z,\ol z)=G_1(y^1)+G_2(y^2)\ ,~~~
h(z,\ol z)=H_1(y^1)+H_2(y^2)\ ,
\end{eqnarray}
where, $G_i$ and $H_i$ ($i=1,2$) are real functions.
Then, (\ref{gh}) can be solved when these functions satisfy
\begin{eqnarray}
 G_1'H_1'= -c_0\ ,~~~ G_2'H_2'=c_0\ ,
\end{eqnarray}
where $c_0$ is a real constant. Note that (\ref{fgh}) implies
\begin{eqnarray}
 e^{-2f}=\frac{H_2'}{G_1'}\ ,
\end{eqnarray}
which makes sense when $H_2'G_1'>0$.

Then, (\ref{hgk}) can be written as
\begin{eqnarray}
e^{i\sigma}\del_z A=\del_z B\ ,
\label{sigdA}
\end{eqnarray}
where we have defined
\begin{eqnarray}
e^{i\sigma}&\equiv&\frac{W+2c_0i}{V}\ ,
\\
W&\equiv& G_1'H_2'+G_2'H_1'
=G_1'H_2'-\frac{c_0^2}{G_1'H_2'}
\ ,
\\
V&\equiv&\sqrt{W^2+4c_0^2}
=G_1'H_2'+\frac{c_0^2}{G_1'H_2'}\ ,
\label{V-def}
\\
A&\equiv& a W+2c_0 k\ ,
\\
B&\equiv& a V\ .
\end{eqnarray}
Some useful identities are
\begin{eqnarray}
G_1'H_2'= c_0\cot\left(\frac{\sigma}{2}\right)\  ,
~~~W=2c_0\cot\sigma\ ,~~~V=\frac{2c_0}{\sin\sigma}\ .
\end{eqnarray}

Note that in the Taylor expansion
of (\ref{sigdA}) with respect to $c_0$,
the leading term is trivially satisfied and the $\cO(c_0)$ term 
reproduces (\ref{aHG}).

Here, we try to solve (\ref{sigdA}) using the following
ansatz:
\begin{eqnarray}
A(y^1,y^2)=A_1(y^1)+A_2(y^2) \ ,~~~
B(y^1,y^2)=B_1(y^1)+B_2(y^2) \ ,
\label{ansatzAB}
\end{eqnarray}
where $A_i$ and $B_i$ are real functions.
Inserting this ansatz into (\ref{sigdA}),
we obtain
\begin{eqnarray}
G_1'H_2'(A_1'-B_1')-\frac{c_0^2}{G_1'H_2'}(A_1'+B_1')
&=&-2c_0 A_2'\ ,
\\
G_1'H_2'(A_2'-B_2')-\frac{c_0^2}{G_1'H_2'}(A_2'+B_2')
&=&2c_0 A_1'\ .
\end{eqnarray}
These equations can be solved when
\begin{eqnarray}
 A_1'+B_1'=2\kappa_2G'_1\ ,~~
 A_1'-B_1'=\frac{2c_0\kappa_1}{G'_1}\ ,~~
 A_2'+B_2'=-2\kappa_1H'_2\ ,~~
 A_2'-B_2'=\frac{2c_0\kappa_2}{H'_2}\ ,
\end{eqnarray}
where $\kappa_i$ ($i=1,2$) are real constants.
Then, we obtain
\begin{eqnarray}
A&=&\kappa_2 G_1-\kappa_1 H_2+c_0(K_1+K_2)+a_0\ ,
\\
B&=&\kappa_2 G_1-\kappa_1 H_2-c_0(K_1+K_2)+b_0\ ,
\end{eqnarray}
where $K_i=K_i(y^i)$ ($i=1,2$) are real functions
satisfying (\ref{AKK}), and $a_0$ and $b_0$ are real constants. 
(We can set $a_0=b_0=0$ by absorbing them in the constant parts of
$G_1$, $H_2$ and $K_i$, but we will keep them for convenience.)

Then, by the definition of $A$ and $B$, we get
\begin{align}
a
=&\
\frac{B}{V}=\frac{B}{2c_0}\sin\sigma\ ,
\\
k
=&\
\frac{1}{2c_0}\left(A-B\frac{W}{V}\right)
=\frac{1}{2c_0}\left(A-B\cos\sigma\right)\ .
\end{align}
Other parameters are obtained by
using (\ref{2nd3}), (\ref{beta+igamma2}), (\ref{r-wtr}), (\ref{dhdg})
and  (\ref{k-ac}):
\begin{align}
c
=&\
\pm\frac{1}{2c_0}\left(W-V\frac{A}{B}\right)
=\pm\left(\cot\sigma-\frac{A}{B\sin\sigma}\right)
\ ,
\\
\Phi
=&\
\log\left(B\del_zl\del_{\ol z}\ol l\right)\ ,
\label{c-sol2}
\\
\varphi_\pm
=&\
\frac{\sigma}{4}+\half{\rm Im}[\log\del_{\ol z}\ol l]\ ,
\\
\beta\pm i\gamma
=&\
\frac{e^{-\frac{\sigma}{2}i}}{\sqrt{2B}}\left(
(\log G_1')'-i(\log H_2')'
\right)\ ,
\\
r-\wt r
=&\
\frac{4}{B}\left((\log G_1')''-(\log H_2')''\right)\ ,
\\
r+\wt r
=&\
\frac{4}{B}\frac{W}{V}\left((\log G_1')''+(\log H_2')''\right)
+\frac{2}{B^3}\left(\kappa_2^2 G_1'^2+\kappa_1^2 H_2'^2
+c_0^2(K_1'^2+K_2'^2)\right)
\nn\\
&\
+\frac{24}{B}\frac{c_0^2}{V^2}
\left(((\log G_1')')^2+((\log H_2')')^2\right)
\nn\\
&\
+\frac{4}{B^2}\left(\frac{W}{V}-1\right)(\kappa_2 G_1''-\kappa_1 H_2'')
-\frac{4c_0}{B^2}\left(\frac{W}{V}+1\right)(K_1''+K_2'')
\nn\\
=&\
\frac{4}{B}\cos\sigma\left((\log G_1')''+(\log H_2')''\right)
+\frac{2}{B^3}\left(\kappa_2^2 G_1'^2+\kappa_1^2 H_2'^2
+c_0^2(K_1'^2+K_2'^2)\right)
\nn\\
&\
+\frac{6}{B}\sin^2\sigma
\left(((\log G_1')')^2+((\log H_2')')^2\right)
\nn\\
&\
-\frac{8}{B^2}\sin^2\left(\frac{\sigma}{2}\right)
(\kappa_2 G_1''-\kappa_1 H_2'')
-\frac{8c_0}{B^2}\sin^2\left(\frac{\sigma}{2}\right)(K_1''+K_2'')
\ .
\label{rwtr-sol2}
\end{align}
One can check that this solution reduces to the one given in the
previous subsection in the $c_0\ra 0$, $\sigma\ra 0$ limit.

\subsection{F-theory configuration: $ISO(1,1)\times SO(6)$}
\label{SO6}

As an explicit example of the solutions obtained in Section \ref{sol1},
let us consider
\begin{eqnarray}
 G_1= y^1\ ,~~~H_2=y^2\ ,~~K_1=\kappa_1 y^1\ ,~~K_2=\kappa_2 y^2\ .
\end{eqnarray}
In this case, (\ref{ak}) gives
\begin{eqnarray}
 a=\kappa_2 y^1-\kappa_1 y^2\ ,~~~
 k=\kappa_1 y^1+\kappa_2 y^2\ ,
\end{eqnarray}
and hence
\begin{eqnarray}
 a+ik = 2(\kappa_2+i\kappa_1) l(z)
\label{a+ik}
\end{eqnarray}
is a holomorphic function of $z$.
This implies that
the complex coupling $\tau\propto i(a\pm ik)$ defined in (\ref{tau})
is holomorphic or anti-holomorphic depending on the chirality
of the unbroken SUSY.
Another immediate consequence is that the combination
$\Phi-\log a\propto {\rm Re}(\log \del_z l)$ is a harmonic function
satisfying the Laplace equation in two dimensions:
\begin{eqnarray}
 \del_z\del_{\ol z} (\Phi-\log a)=0\ .
\label{harmonic}
\end{eqnarray}
Actually, in this case, (\ref{c-sol1})--(\ref{rwtr-sol1})
imply $\beta=\gamma=0$ and $r=\wt r$,
and the symmetry $SO(3)\times SO(3)$ is enhanced to $SO(6)$.

Though this solution is a special solution, the properties
(\ref{a+ik}) and (\ref{harmonic}) hold for general solutions
with $ISO(1,1)\times SO(6)$ symmetry. In fact, it is not difficult to
find general solutions for the cases with
$ISO(1,1)\times SO(6)$ symmetry.
If we impose $\beta=\gamma=0$,  (\ref{beta+igamma2})
implies $g={\rm constant}$ or $f={\rm constant}$.
However, in order to have non-singular solution of (\ref{fdelg}), $g$
cannot be a constant. Then, $f$ has to be a constant and
(\ref{gf}) implies that
 $g$ is a harmonic function satisfying
$\del_z\del_{\ol z} g=0$.
Then, (\ref{fdelg}) and (\ref{afgk}) imply that
\begin{eqnarray}
\Phi-\log a-4 i\varphi_\pm
\end{eqnarray}
and $a+ik$ are holomorphic functions.
Therefore the general solution can be written as
\begin{eqnarray}
&& a=\ol l(\ol z)+l(z)\ ,~~k=i(\ol l(\ol z)-l(z))\ ,
\\
&& \Phi=\log(\ol l(\ol z)+l(z))+\ol m(\ol z)+m(z)\ ,~~
\varphi_\pm=- \frac{i}{4}(\ol m(\ol z)-m(z))
\end{eqnarray}
with holomorphic functions $l(z)$ and $m(z)$.

Note that all the parameters in the Lagrangian
are invariant under a constant shift of the imaginary part of
$m(z)$, but $\varphi_\pm$ is shifted.
This shift induces a shift of $\xi$ in (\ref{ep0-2}) through
(\ref{inteps0}) and hence we can choose any value for $\xi$ without
changing the action. Therefore, as explained below (\ref{ep0-2}),
the number of preserved SUSY is enhanced to 8.

This solution corresponds to the D3-brane probes in the F-theory
configurations (background with 7-branes in type IIB string theory)
\cite{Harvey:2007ab,Buchbinder:2007ar,Harvey:2008zz}
and the equation (\ref{harmonic}) is interpreted
as the Einstein equation \cite{Greene:1989ya}.
To see this explicitly, note that the type IIB supergravity action for
the dilaton $\phi$, RR 0-form $C_0$ and gravity is given by
\begin{eqnarray}
 S_{\rm IIB}=\frac{1}{2\kappa_{10}^2}\int d^{10}x\sqrt{-g}
\left(R-\frac{\del_\mu\ol\tau\del^\mu\tau}{2({\rm Im}\,\tau)^2}
+\cdots\right)\ ,
\end{eqnarray}
where $\tau\equiv C_0+ie^{-\phi}$. Assuming that $\tau$ only depends
on $x^2$ and $x^3$, and using the ansatz (\ref{metric}) for the metric,
the Einstein equation and the equation of motion for $\tau$ become
\begin{eqnarray}
-\del_z\del_{\ol z}\Phi+\frac{\del_z\tau\del_{\ol z}\ol\tau
+\del_{\ol z}\tau\del_{z}\ol\tau}{(\tau-\ol\tau)^2}&=&0\ ,
\label{PhiEOM}
\\
\del_z\tau\del_z\ol\tau&=&0\ ,
\\
 \del_z\del_{\ol z}\tau+2\frac{\del_z\tau\del_{\ol z}\tau}
{\ol\tau-\tau}&=&0\ .
\end{eqnarray}
Then, it is easy to see that $\tau=\tau(z)$ (a holomorphic function of
$z$) solves the equations of motion for $\tau$
and then (\ref{PhiEOM}) becomes
\begin{eqnarray}
 \del_z\del_{\ol z}(\Phi-\log{\rm Im}\tau)=0\ ,
\end{eqnarray}
which agrees with (\ref{harmonic}) with the identification
$\tau\propto i(a+ik)$.

\subsection{Gaiotto-Witten solution: $ISO(1,2)\times SO(3)\times SO(3)$}
\label{GW}

The supersymmetric Janus configuration found by Gaiotto-Witten in
\cite{Gaiotto:2008sd} can be obtained as a special solution of
the solutions obtained in Section \ref{sol2}.
It is a solution with $ISO(1,2)\times SO(3)\times SO(3)$ symmetry
and all the deformation parameters depend only on one coordinate $x^3$.

To see this, let us consider the case with
$\kappa_i=0$, $K_i=0$ ($i=1,2$) and $H_2=y^2$.
In this case, $\Phi$ is a harmonic function
and, as discussed in Section \ref{more}, we may have solutions
with both $\ol\epsilon_+$ and $\ol\epsilon_-$ being non-zero.
To obtain such solutions, we should make sure
that the parameters in the Lagrangian are consistent with
the SUSY conditions for both $\ol\epsilon_+$ and $\ol\epsilon_-$
simultaneously. To this end, we choose
\begin{eqnarray}
 c_0=\pm \frac{1}{D}\ ,~~\sigma=\pm 2\psi
\end{eqnarray}
for the solution in (\ref{c-sol2})$\sim$(\ref{rwtr-sol2}).
Then, we have the relation $\cot\psi=D G_1'(y^1)$ and
\begin{eqnarray}
a&=&\frac{b_0D}{2}\sin(2\psi)\ ,
\\
c&=&
\cot(2\psi)-\frac{a_0}{b_0\sin(2\psi)}\ ,
\\
\Phi&=&\log\left(b_0\del_zl\del_{\ol z}\ol l\right)\ ,
\\
\varphi_\pm
&=&\pm\frac{\psi}{2}+\half{\rm Im}[\log\del_{\ol z}\ol l]\ ,
\\
 \beta\pm i\gamma&=&\frac{e^{\mp i\psi}}{\sqrt{2b_0}}(\log \cot\psi)'\ ,
\\
r-\wt r
&=&\frac{4}{b_0}(\log \cot\psi)''\ ,
\\
r+\wt r
&=&
\frac{4}{b_0}\cos(2\psi)(\log\cot\psi)''
+\frac{6}{b_0}\sin^2(2\psi)((\log\cot\psi)')^2
\ .
\end{eqnarray}
This agrees with the supersymmetric Janus solution in 
\cite{Gaiotto:2008sd} when
$l(z)=-i z/\sqrt{2}$ and $b_0=2$. 
In this configuration, 8 supersymmetries are preserved.

\section{Other examples}
\label{S04}

\subsection{$ISO(1,1)\times SO(2)\times SO(4)$}
\label{Ex24}

Similar to the case with $ISO(1,1)\times SO(3)\times SO(3)$ symmetry
considered in the previous section,
we expect to have SUSY preserving configurations with
$ISO(1,1)\times SO(2)\times SO(4)$ symmetry, as it can be realized
in a D3-D3(-D7) system in the following table:
\begin{eqnarray}
\begin{tabular}{c|ccccccccccccc}
&0&1&2&3&4&5&6&7&8&9\\
\hline
D3&o&o&o&o\\
D3&o&o&&&o&o\\
D7&o&o&&&o&o&o&o&o&o
\end{tabular}
\end{eqnarray}
In this case, we regard the D3-brane extended along $x^{0\sim 3}$
directions as a probe embedded in the supergravity background
corresponding to the other D3 and D7-branes. The 4 dimensional
gauge theory with varying couplings is realized on the world-volume of
the probe D3-brane.

\subsubsection{Ansatz for deformation}

First, we set the metric and other deformation parameters consistent
with the global symmetry. The ansatz for the metric (\ref{metric})
is the same as that used in Section \ref{33ansatz}:
\beal{
ds^2 = \eta_{\alpha \beta} dx^\alpha  dx^\beta + e^{\Phi (x^i)}
\delta_{ij} dx^i dx^j + \delta_{ab} dx^a dx^b + \delta_{pq}  dx^p dx^q\ ,
}
where the indices are $\alpha, \beta=0,1$; 
$i,j=2,3$; $a, b=4,5$ and $p, q=6,7,8,9$. The non-trivial
components of the spin connection are given in (\ref{spinconn}).
The parameters $M$, $d^{IJA}$ and $m^{AB}$
consistent with the symmetry are of the form
\beal{
M= 6\left(m_{01 i}\Gamma^{01i}+m_{i45}\Gamma^{i45}\right)
\equiv \alpha_i \G^i \G^{01} + \beta_i \G^i \G^{45}\ ,
}
\beal{
d^{iab}=-d^{aib} = v^i \epsilon^{ab}\ ,~~~
m^{ab} = r \delta^{ab},~~~
m^{pq}=\wt{r} \delta^{pq}\ ,
}
where $\epsilon^{ab}$ is the epsilon tensor
satisfying $\epsilon^{45}=-\epsilon^{54}=1$.
Here, $\alpha_i$, $\beta_i$, $v^i$, $r$ and $\wt r$ are functions of $x^i$.

\subsubsection{Solutions of the SUSY conditions}

First, let us apply the above ansatz to the first order equations
\eqref{SUSYcond1-e}--\eqref{Dmueq2}.
It is easy to show that all the components of $e^{IJK}$
vanish. The components of $e^{IJK}$ that are allowed by the symmetry are
\begin{eqnarray}
e^{01i}&=&  a^{-1} \p_j (ac) \epsilon^{ji} -4 \alpha^i ,\\
e^{i45}&=& 4 \beta^i + 2 v^i . 
\end{eqnarray}
Equation \eqref{SUSYcond1-e} for $(I,J)=(6,7)$ implies
\beal{
0 =\eb \, e^{I' J' K'} \G_{K'} \G_{I'J'}
 =6\, \eb (e^{01i} \G_{01i}+ e^{i45} \G_{i45} )\ .
}
From this equation, together with (\ref{SUSYcond1-e}) for $(I,J)=(4,6)$,
we obtain
\beal{
\eb\, e^{01i} \G_{i}=\eb\, e^{i45} \G_{i}=0\ .
}
Because $\Gamma^{\hat 2\hat 3}$ does not have a real eigenvalue, we
conclude that $e^{01i}=e^{i45}=0$ to have non-zero $\ol\epsilon$.
Therefore, we obtain
\beal{
 -2 \beta^i =v^i\ ,~~~ 
 a^{-1} \p_j (ac)\epsilon^{ji} = 4 \alpha^i\ .
\label{beta-alpha}
} 

In this case, (\ref{am3}) gives
\begin{eqnarray}
0=\ol\epsilon\,\left(
\Gamma^{i}q_i-4\alpha_i\Gamma^{01i}
\right)\ ,
\label{qal}
\end{eqnarray}
where $q_i\equiv \del_i\log a$.
When we decompose $\ol\epsilon$ as
$\ol\epsilon=\ol\epsilon_++\ol\epsilon_-$ with
$\ol\epsilon_\pm\Gamma^{01}=\pm\ol\epsilon_\pm$,
(\ref{qal}) becomes
\begin{eqnarray}
0=\ol\epsilon_+(q_i-4\alpha_i)\Gamma^{i}
+\ol\epsilon_-(q_i+4\alpha_i)\Gamma^{i}\ .
\end{eqnarray}
This is satisfied if and only if
\begin{eqnarray}
q_i=\pm 4\alpha_i~~~\mbox{and}~~~
\ol\epsilon_\mp=0\ ,
\label{case2-24}
\end{eqnarray}
or
\begin{eqnarray}
q_i=\alpha_i=0\ .
\label{case1-24}
\end{eqnarray}
Note that the second equation of (\ref{beta-alpha}) implies
that the complex coupling (\ref{tau}) satisfies
\footnote{The notation for complex coordinates is given in Appendix
\ref{cpxcoord}.}
\begin{eqnarray}
 \del_z\tau=4\pi a\,i (4\alpha_z+q_z)\ ,~~~
 \del_{\ol z}\tau=4\pi a\,i (-4\alpha_{\ol z}+q_{\ol z})\ .
\end{eqnarray}
From this we find that the complex coupling $\tau$ is
holomorphic or
anti-holomorphic when $q_i=4\alpha_i$ or $q_i=-4\alpha_i$, respectively.
This is the same situation as that observed in section \ref{SO6}.
For the case with (\ref{case1-24}), the gauge coupling and theta parameter
are constant.

From (\ref{cA}), we obtain the components
\begin{eqnarray}
&&\cA_0= -\frac{1}{4}\left(
F\Gamma_0+
4\alpha_i\Gamma^{1i}
\right)
\ ,
~~~
\cA_1= -\frac{1}{4}\left(
F\Gamma_1-
4\alpha_i\Gamma^{0i}
\right)
\ ,
\nn\\
&&\cA_i= -\frac{1}{4}\left(
F\Gamma_i+
4\alpha_i\Gamma^{01}+4\beta_i\Gamma^{45}
-\epsilon_i^{~j}\del_j\Phi\Gamma^{\hat 2\hat 3}
\right)
\ .
\end{eqnarray}
We find that (\ref{Falpha}) is also valid in this case. Then,
(\ref{Beq2}) implies
\begin{eqnarray}
\eb\ol{B}^a= - 4\,\eb\,(\alpha_i \G^{i01} - \beta_i \G^{i45})\G^a\ ,~~~
\eb\ol{B}^p= - 4\,\eb\, \alpha_i \G^{i01}\G^p\ ,
\label{B-24}
\end{eqnarray}
and (\ref{Dmueq2}) with $\mu=i=2,3$ can be written as
\begin{eqnarray}
\del_i\ol\epsilon=
\ol\epsilon\left(-\frac{1}{4}\del_j\Phi\Gamma^{j}_{~i}+
\alpha_j\Gamma^{01}\Gamma^j_{~i}-\beta_i\Gamma^{45}
\right)\ .
\label{delepsilon-24}
\end{eqnarray}
The integrability condition (\ref{eF}) for (\ref{delepsilon-24}) is
equivalent to
\begin{eqnarray}
 0=\ol\epsilon
\left(-\frac{1}{4}g^{ij}\del_i\del_j\Phi
+g^{ij}\del_i\alpha_j\Gamma^{01}
+\epsilon^{ij}\del_i\beta_j\Gamma^{\hat2\hat3 45}
\right)\ .
\label{integrability-24}
\end{eqnarray}

In order to solve this equation, we decompose $\eb$ into
the eigenspaces of $\Gamma^{01}$ and $\Gamma^{\hat2\hat3 45}$ as
\begin{eqnarray}
 \eb=\eb_{+}^{+}+\eb_{+}^{-}+\eb_{-}^{+}+\eb_{-}^{-}\ ,
\end{eqnarray}
where $\eb^s_t$ ($s=\pm$, $t=\pm$) satisfy
\begin{eqnarray}
\eb^s_t\,\Gamma^{01} =t\,\eb^s_t\ ,~~~
\eb^s_t\,\Gamma^{\hat2\hat345} =s\,\eb^s_t\ ,~~~
\eb^s_t\,\Gamma^{(10)}= \eb^s_t\ .
\label{ebxieta}
\end{eqnarray}
Equation (\ref{integrability-24}) implies that $\eb^s_t$
can be non-zero only if
\begin{eqnarray}
0= -\frac{1}{4}g^{ij}\del_i\del_j\Phi
+t g^{ij}\del_i\alpha_j
+s \epsilon^{ij}\del_i\beta_j
\label{int-24-2}
\end{eqnarray}
is satisfied.
This equation can be solved when there exists a function
$\varphi^s_t$ satisfying
\begin{eqnarray}
 -\frac{1}{4}\del_j\Phi+t\alpha_j+s\epsilon_j^{~i}\beta_i
=\epsilon_j^{~i}\del_i\varphi_t^s\ .
\label{varphi-24}
\end{eqnarray}
Then, the solution of (\ref{delepsilon-24}) is given by
\begin{eqnarray}
\eb^s_t = \eb^{0s}_t e^{\varphi^s_t \Gamma^{\hat2\hat3}}\ ,
\end{eqnarray}
where $\eb_t^{0s}$ is a constant spinor satisfying
the conditions in (\ref{ebxieta}).

Then we distinguish the following four cases:
\begin{itemize}
\item (C1):
$\alpha_j\ne 0$ and $\epsilon^{ij}\del_i\beta_j\ne0$. \\
Equations (\ref{case2-24}) and (\ref{int-24-2}) imply
that only one combination of the signs $(t,s)$ can have
non-zero $\eb^s_t$.
\item (C2):
$\alpha_j\ne 0$ and $\epsilon^{ij}\del_i\beta_j=0$. \\
Equation (\ref{case2-24}) implies that only one sign for $t$ is allowed,
but both $\eb^+_t$ and $\eb^-_t$ can be non-zero.
In this case (\ref{case2-24}) and (\ref{int-24-2}) imply
\begin{eqnarray}
g^{ij} \del_i\del_j(\Phi-\log a)=0\ ,
\label{Phi-loga}
\end{eqnarray}
and there exist functions
$\varphi_1$ and $\varphi_2$ satisfying
\begin{eqnarray}
-\frac{1}{4}\del_j(\Phi-\log a)=\epsilon_j^{~i}\del_i\varphi_1
\ ,~~~
\beta_j=\del_j\varphi_2\ .
\label{varphi1-2}
\end{eqnarray}
Then, $\varphi_t^s\equiv \varphi_1+s\varphi_2$ satisfies (\ref{varphi-24}).
\item (C3):
$\alpha_j=0$ and $\epsilon^{ij}\del_i\beta_j\ne 0$. \\
Equation (\ref{int-24-2}) implies that only one sign for $s$ is allowed,
but both $\eb^s_+$ and $\eb^s_-$ can be non-zero.
\item (C4):
$\alpha_j=0$ and $\epsilon^{ij}\del_i\beta_j= 0$. \\
All the sixteen components of $\eb$ can be non-zero.
Equation (\ref{int-24-2}) implies $g^{ij} \del_i\del_j\Phi=0$.
\end{itemize}
  
Next, let us consider the second order equation  (\ref{SUSYcond2}).
In this case, it can be written as
\begin{eqnarray}
D_i(\eb\ol B^a)\G^i&=&\eb\left(
-2(D_iv^i+q_iv^i)\epsilon^{ab}\G_b-r\G^a
-\ol B^a \left(M+\half\G^iq_i\right)
\right)\ ,
\label{2nd-24-1}
\\
D_i(\eb\ol B^p)\G^i&=&\eb\left(
-\wt r\G^a
-\ol B^p \left(M+\half\G^iq_i\right)
\right)\ .
\label{2nd-24-2}
\end{eqnarray}
Inserting (\ref{B-24}) into these equations and using
(\ref{delepsilon-24}), we obtain
\begin{eqnarray}
0&=&\eb\left(r+\half g^{ij} q_iq_j+g^{ij}\del_i q_j-8g^{ij}\beta_i\beta_j
+4\del_i\beta_j\epsilon^{ij}\G^{\hat2\hat3 45}
\right)\ ,
\label{r-24}
\\
\wt r&=&-\half g^{ij} q_iq_j-g^{ij}\del_i q_j
\ ,
\label{wtr-24}
\end{eqnarray}
where we have used the relation in \eqref{beta-alpha},
\begin{eqnarray}
0= \eb\left(q_i+4\alpha_i\G^{01}\right)\ ,
\end{eqnarray}
that is valid for both cases (\ref{case2-24}) and (\ref{case1-24}). Then, from 
(\ref{r-24}) we obtain
\begin{eqnarray}
 r=-\half g^{ij} q_iq_j-g^{ij}\del_i q_j+8g^{ij}\beta_i\beta_j
-4s\del_i\beta_j\epsilon^{ij}
\ ,
\label{r-24-2}
\end{eqnarray}
where we have used (\ref{ebxieta}) in order to have non-zero $\eb_t^s$.

In summary, we can construct a generic solution of the SUSY conditions
by the following steps. First, pick a holomorphic
 (or anti-holomorphic) function $\tau(z)$ such that ${\rm Im}\,\tau >0$.
Then, $a$ and $c$ are obtained from (\ref{tau}) and
$\alpha_i$ is given by (\ref{case2-24}). Next, choose arbitrary
real functions $\Phi$ and $\varphi^s_t$. Then,
$\beta_i$ is determined by (\ref{varphi-24}). If one wants to
find a solution with $\epsilon^{ij}\del_i\beta_j=0$,
(the case (C2) above), $\Phi$ is determined by solving (\ref{Phi-loga}).
More explicitly, the solutions of (\ref{Phi-loga}) are obtained
by choosing a holomorphic function $l(z)$ and setting
\begin{eqnarray}
 \Phi=\log a + l+\ol l\ .
\end{eqnarray}
$\varphi_1$ in (\ref{varphi1-2}) is given by
\begin{eqnarray}
 \varphi_1=\frac{i}{4}(l-\ol l) \ .
\end{eqnarray}
Other parameters $v$, $r$ and $\wt r$ are obtained
by solving equations (\ref{beta-alpha}), (\ref{r-24-2}) and
(\ref{wtr-24}), respectively.

The number of unbroken SUSY is 4 for the generic case (C1),
8 for the cases (C2) and (C3), and 16 for (C4).
Case (C4) is a trivial solution that is related to the undeformed
${\cal N}=4$ SYM by a coordinate transformation and a field
redefinition. In fact, because $\Phi$ is harmonic, it can be eliminated
by a conformal transformation on the $z$-plane. $\beta_i$ is a pure
gauge configuration and it can be eliminated by a local $SO(6)$
rotation (see Appendix \ref{SO6tr}).
One can guess that the cases (C1), (C2) and (C3) correspond to the
D3-D3-D7, D3-D7, D3-D3 systems, respectively \cite{preparation}.

\subsection{Time dependent solutions}
\label{timedep}

\subsubsection{$ISO(3)$}
\label{ISO3}

Let us consider the cases in which the couplings depend on time.
First, as a trial, let us assume that spatial translational
and rotational symmetry $ISO(3)$ are preserved. It turns out that
the time dependent solutions of the SUSY conditions with this symmetry
can always be mapped to a system with a constant gauge coupling and theta
parameter by general coordinate transformations and field
redefinitions.\footnote{We have not been able to exclude the possibility
that mass parameters have non-trivial time dependence, despite the gauge
coupling and theta parameter are constant.
}

Note that the metric (\ref{metric}) can be chosen to be flat.
This is because the general form of metric preserving $ISO(3)$ symmetry
(flat FLRW metric) can be written as
\begin{eqnarray}
 ds^2=e^{\Phi(\eta)}(-d\eta^2+\delta_{ij}dx^idx^j)+\delta_{AB}dx^Adx^B\ ,
\label{FRWmetric}
\end{eqnarray}
where $i,j=1,2,3$ and $\eta$ is the conformal time defined by
 $d\eta=e^{-\Phi/2}dt$. The overall factor  $e^\Phi$ in the
 4-dimensional metric can be eliminated by the Weyl 
 transformation\footnote{See Appendix \ref{Weyl}.}
 ($g_{\m \n} \rightarrow e^{-\Phi}g_{\m \n}$)
without loss of generality. 

Then, it is particularly easy to show that the gauge coupling
cannot depend on time to preserve SUSY. To see this,
note that (\ref{am2}) may be written as
\beal{
\eb \big(\p_0 \log a \big) =
\ol\epsilon\,\bigg(\frac{1}{36}e^{IJK}\Gamma_{IJK}\G_0 
-\left(
\frac{1}{8}e_{\mu JK}-6m_{\mu JK}\right)\Gamma^{\mu JK}\G_0
-2 M \G_0
\bigg)\ .
}
Because $\Gamma_{IJK}\Gamma_0$ is an anti-symmetric matrix
for all $I,J,K=0,\dots,9$, the right hand side of this equation is
$\eb$ times a real anti-symmetric matrix. However, since
a real anti-symmetric matrix cannot have a real non-zero eigenvalue,
the only possibility is that both the left and right hand sides are
zero:
\beal{
\p_0 a =0\ , ~~~
\ol\epsilon\,\bigg(\frac{1}{36}e^{IJK}\Gamma_{IJK} -\left(
\frac{1}{8}e_{\mu JK}-6m_{\mu JK}\right)\Gamma^{\mu JK}
-2 M \bigg) =0\ .
\label{del0a}
}
Therefore, the gauge coupling has to be time independent.

As explained in Appendix \ref{e-ISO3}, all the components of $e^{IJK}$
vanish for the cases with $ISO(3)$ symmetry.
Because non-trivial components of $m^{IJK}$ are $m^{ABC}$, $m^{0AB}$ and
$m^{123}$, the second equation of \eqref{del0a} becomes
\beal{
\eb (-12 \, m^{123} )= \eb \, m_{ABC}\G^{ABC}\G^{123}\ ,
}
which means that $m^{123}$ is proportional to the eigenvalue
of $m_{ABC}\G^{ABC}\G^{123}$. However, since
$m_{ABC}\G^{ABC}\G^{123}$ is a real anti-symmetric matrix,
the eigenvalue cannot take a non-zero real value. The only
possibility is
\begin{eqnarray}
m^{123}=0\ ,~~~\eb\, m_{ABC}\Gamma^{ABC}=0\ .
\label{m-ISO3}
\end{eqnarray}
Then, (\ref{delac}) implies $\del_0(ac)=0$ and, as a consequence, the
theta parameter \eqref{couplings} is also a constant. 

Furthermore, (\ref{Dmueq2}) with $\mu=1,2,3$ and (\ref{cA}) implies
 $0=\eb\cA_i\propto\eb F\Gamma_i$ and we have
\begin{eqnarray}
\del_0\eb=\eb \cA_0=-3\,\eb\, m_{0 AB}\Gamma^{AB}\ .
\end{eqnarray}
We can set $m_{0AB}=0$ by the local $SO(6)$
transformation\footnote{See Appendix \ref{SO6tr}.}
and then the SUSY parameter $\eb$ is also constant.

\subsubsection{$ISO(2)\times SO(6)$}

To get a solution with time dependent gauge coupling,
we consider the cases in which the couplings can depend on $x^0$ and $x^1$.
We impose the $ISO(2)$ symmetry that acts on the $x^{2,3}$ directions
and $SO(6)_R$ symmetry to simplify the analysis.
The ansatz for the metric is
\begin{eqnarray}
 ds^2=e^{\Phi(x^\alpha)}\eta_{\alpha\beta}dx^\alpha dx^\beta
+\delta_{ij}dx^idx^j +\delta_{AB}dx^Adx^B\ ,
\end{eqnarray}
where $\alpha,\beta=0,1$; $i,j=2,3$ and $A,B=4,5,\cdots,9$.
The non-trivial components of the spin connection are
\begin{eqnarray}
\omega_{0 \hat{0} \hat{1} }= -\f{\p_1 \Phi}{2}\ ,
~~~\omega_{1 \hat{0} \hat{1}}=-\f{\p_0 \Phi}{2}\ .
\end{eqnarray}
Because of the $ISO(2)\times SO(6)$ symmetry, all the components of
$d^{IJK}$ are zero and the form of $M$ consistent with the symmetry is
\begin{eqnarray}
M=6 m_{\alpha 23}\Gamma^{\alpha 23}
 \equiv \alpha_\alpha \Gamma^{\alpha 23}\ .
\end{eqnarray}
Then,
$e^{\mu A I}$ ($\mu=0,\cdots,3$ and $A=4,\cdots,9$)
are all zero and we can use the argument given in Appendix \ref{e-mixed}
to conclude that all the components of $e^{IJK}$ are zero. In addition, 
(\ref{delac}) and (\ref{delm}) imply
\beal{
a^{-1} \p_0 (ac)
 =-4 \alpha_1\ ,
~~~
a^{-1} \p_1 (ac)
 =-4 \alpha_0\ ,
~~~
g^{\alpha\beta} \p_\alpha (a \alpha_\beta)=0\ ,
\label{acalpha}
}
and \eqref{am3} becomes
\beal{
0=\eb\, (  q_\alpha \G^\alpha  -4 \alpha_\alpha\Gamma^{\alpha 23})\ ,
\label{1sto-ISO2-1}
}
where $q_\alpha = \p_\alpha \log a$.
Equation (\ref{1sto-ISO2-1}) is equivalent to
\begin{eqnarray}
0=
\eb_\pm \left((q_0\pm q_1 )-4 (\alpha_0\pm \alpha_1)\Gamma^{23}\right)
\ ,
\end{eqnarray}
where $\eb$ is decomposed as $\eb =\eb_{+} + \eb_{-}$
with $\eb_\pm\Gamma^{\hat0\hat1}=\pm\eb_\pm$. Since $\Gamma^{23}$ does
not have a real eigenvalue, this equation implies
\begin{eqnarray}
 q_0=\mp q_1\ ,~~~\alpha_0=\mp\alpha_1~~~
\mbox{and}~~~\eb_\mp=0\ ,
\label{qalpha}
\end{eqnarray}
or a trivial solution;
\begin{eqnarray}
q_\alpha=\alpha_\alpha=0\ . 
\end{eqnarray}

In the following we focus on the case in (\ref{qalpha}).
Equations (\ref{qalpha}) and (\ref{acalpha}) imply
\begin{eqnarray}
 (\del_0\pm\del_1)a=0\ ,~~~
 (\del_0\pm\del_1)(ac)=0\ ,~~~
 (\del_0\pm\del_1)\alpha_\alpha=0\ ,
\label{delalpha}
\end{eqnarray}
and the solution can be written as
\begin{eqnarray}
\alpha_0=\mp\alpha_1=h_\mp'(x^\mp)\ ,~~~
 ac = k_\mp(x^\mp)\ ,~~~
a=\pm\frac{k_\mp'(x^\mp)}{4 h_\mp'(x^\mp)}\ ,
\label{sol-ISO2}
\end{eqnarray}
or
\begin{eqnarray}
\alpha_\alpha=0\ ,~~~c=0~~~\mbox{and}~~~a=a_\mp(x^\mp)\ ,
\label{c=0}
\end{eqnarray}
where $h_\mp$, $k_\mp$ and $a_\mp$
are arbitrary real functions
of $x^\mp\equiv x^0\mp x^1$
and prime denotes the derivative.\footnote{
The solution (\ref{c=0}) was discussed in \cite{Kim:2008dj}.
}

From \eqref{cA}, $\cA_\mu$ is given by
\begin{eqnarray}
\cA_0&=&
-\frac{1}{4}\left(F\Gamma_0+4\alpha_{0}\Gamma^{23}
+\del_1\Phi\Gamma^{\hat 0\hat 1}
\right)
\ ,\\
\cA_1&=&
-\frac{1}{4}\left(F\Gamma_1+4\alpha_1\Gamma^{23}
+\del_0\Phi\Gamma^{\hat0\hat1}
\right)
\ ,\\
\cA_i&=&
-\frac{1}{4}\left(F\Gamma_i+
\epsilon_{ij} 4\alpha_{\alpha}
\Gamma^{j \alpha}
\right)
\ .
\end{eqnarray}
For $\mu=i$, the condition \eqref{Dmueq2} is $0=\del_i\eb=\eb\cA_i$ and implies
\begin{eqnarray}
 \eb\,F=-4\, \eb\,\alpha_\alpha\Gamma^{\alpha23}=0\ ,
\end{eqnarray}
where we have used (\ref{qalpha}).
Then, from (\ref{Beq2}) and (\ref{Dmueq2}) we obtain, respectively,
\begin{eqnarray}
 \eb\,\ol{B}^A=0
\label{B-ISO2}
\ ,
\end{eqnarray}
and
\begin{eqnarray}
\del_0\eb=-\eb
\left(\alpha_0\Gamma^{23}+\frac{1}{4}\del_1\Phi\Gamma^{\hat0\hat1}
\right)\ ,
~~~
\del_1\eb=-\eb
\left(\alpha_1\Gamma^{23}+\frac{1}{4}\del_0 \Phi\Gamma^{\hat0\hat1}
\right)\ .
\label{delep-ISO2}
\end{eqnarray}

The integrability condition (\ref{eF}) for (\ref{delep-ISO2}) is
\begin{eqnarray}
 0=\eb
\left((\del_1\alpha_0-\del_0\alpha_1)\Gamma^{23}
+\frac{1}{4}\eta^{\alpha\beta}\del_\alpha\del_\beta\Phi
\Gamma^{\hat0\hat1}
\right)\ ,
\end{eqnarray}
which implies
\begin{eqnarray}
\del_1\alpha_0-\del_0\alpha_1=0\ ,~~~ 
g^{\alpha\beta}\del_\alpha\del_\beta\Phi=0\ .
\label{int-ISO2}
\end{eqnarray}
The solutions (\ref{sol-ISO2}) and (\ref{c=0})
satisfy the first equation of (\ref{int-ISO2})
and the general solution of the second equation is
\begin{eqnarray}
 \Phi=f_+(x^+)+f_-(x^-)\ ,
\end{eqnarray}
where $f_\pm$ are arbitrary real functions of $x^\pm$.
Then, the equations (\ref{delep-ISO2}) can be integrated as
\beal{
\eb_{\pm}=\eb^0_{\pm} \,
 e^{ \mp \f{1}{4}( f_+ - f_- )-h_\mp\G^{23} },
}
where $\eb^0_\pm$ is a constant spinor satisfying
$\eb_\pm\Gamma^{\hat0\hat1}=\pm\eb_\pm$.
Using the above results, the second order equation (\ref{SUSYcond2})
becomes simply
\begin{eqnarray}
0= \eb\,m^{AB}\Gamma_B\ .
\end{eqnarray}
Multiplying this equation by $m_{AC}\Gamma^C$, we find
$m_{AB}=0$.

\section{Conclusion and outlook}
\label{conclusion}

We have studied the deformations of $\mathcal{N}=4$ SYM that preserve
SUSY with space-time dependent couplings. 
A lot of explicit solutions of the SUSY conditions have been found.
For example, we have found wide classes of solutions
for the cases with $ISO(1,1) \times SO(3) \times SO(3)$
symmetry and the general solutions for the cases
with $ISO(1,1)\times SO(6)$ and $ISO(1,1) \times SO(2)\times SO(4)$
symmetries. Time dependent cases with $ISO(3)$ and $ISO(2)\times SO(6)$
symmetries have also been analyzed.

As we mentioned in the introduction, it is commonly faced situation
that gauge couplings are not constant when one tries to engineer
a gauge theory using D-branes in string theory.
Therefore, it is natural to ask whether there are string theory
realizations of the solutions that we found in this paper.
We hope to address this question in our forthcoming paper \cite{preparation}.

Our analysis in this paper is classical
and it would be important to take into account the quantum effects.
Since these theories preserve SUSY, we may be able to use
some techniques, such as localization, to calculate some physical
quantities exactly.
Since our system can be realized as D3-branes embedded in type IIB
string theory, we may be able to study S-duality and holographic dual.
It would also be interesting to consider how the
BPS solitonic objects, such as dyons or instantons, behave
when the couplings depend on space-time.

As mentioned in Section \ref{NotationAnsatz},
our ansatz for the action in (\ref{action1}) is not completely general,
even if we restrict the deformation to be of dimension less than 4.
As an obvious extension of the analysis, one may try to include
all the terms that are compatible with renormalizability.
It would also be interesting to consider the case with $U(N)$
gauge group and include the terms like
$\tr F_{IJ}$, $\tr A_A$.
The $U(1)$ part will be important to consider the configurations
of D3-branes, when the system is embedded in string theory.
Since our strategy should work for arbitrary supersymmetric theory,
further generalization would also be possible.

\section*{Acknowledgements}

We thank Takahiro Nishinaka, Shuichi Yokoyama and Ki-Myeong Lee
for discussion.
We appreciate useful discussions during the workshops ``Geometry,
Duality and Strings'' YITP-X-16-11, ``Strings and Fields''
YITP-W-17-08 and the KIAS-YITP Joint Workshop 2017 ``Strings, Gravity 
and Cosmology'' YITP-W-17-12, held at the Yukawa Institute for Theoretical 
Physics, Kyoto University.
J.J.F.-M. gratefully acknowledges the support of JSPS (Postdoctoral
Fellowship) and the Fundaci\'on S\'eneca/Universidad de Murcia (Programa
Saavedra Fajardo).
The work of S.S. was supported by JSPS KAKENHI (Grant-in-Aid for
Scientific Research (C)) Grant Number JP16K05324.
The work of J.J.F.-M. and S.S. was also supported by JSPS KAKENHI
(Grant-in-Aid for JSPS Fellows) Grant Number JP16F16741.

\appendix

\section{Complex coordinates}
\label{cpxcoord}

We use complex coordinates to parametrize the 2-3 plane. Our
convention used in Sections \ref{Ex33} and \ref{Ex24}
is as follows:
\begin{eqnarray}
z=\frac{1}{\sqrt{2}}(x^2+ix^3)\ ,~~~
\ol z=\frac{1}{\sqrt{2}}(x^2-ix^3)\ ,
\end{eqnarray}
\begin{eqnarray}
\del_z=\frac{1}{\sqrt{2}}(\del_2-i\del_3) \ ,~~~
\del_{\ol z}=\frac{1}{\sqrt{2}}(\del_2+i\del_3) \ ,
\end{eqnarray}
\begin{eqnarray}
g_{z\ol z}=g_{\ol z z}= e^\Phi\ ,~~~
g^{z\ol z}=g^{\ol z z}= e^{-\Phi}\ ,
\end{eqnarray}
\begin{eqnarray}
 \epsilon^{z\ol z}=- \epsilon^{\ol z z}=-ie^{-\Phi}\ ,~~~
 \epsilon_{z\ol z}=- \epsilon_{\ol z z}=ie^{\Phi}\ ,~~~
 \epsilon^{z}_{~z}= \epsilon_{\ol z}^{~\ol z}=-i\ ,~~~
 \epsilon^{\ol z}_{~\ol z}= \epsilon_{z}^{~z}=i\ .
\end{eqnarray}
\begin{eqnarray}
&& \del_z\tau=4\pi a\,i (4\alpha_z+q_z)\ ,~~~
 \del_{\ol z}\tau=4\pi a\,i (-4\alpha_{\ol z}+q_{\ol z})\ ,
\nn\\
&& \del_z\ol\tau=4\pi a\,i (4\alpha_z-q_z)\ ,~~~
 \del_{\ol z}\ol\tau=4\pi a\,i (-4\alpha_{\ol z}-q_{\ol z})\ .
\label{deltau}
\end{eqnarray}
\begin{eqnarray}
4\alpha_z=-i a^{-1}\del_z(ac)\ ,~~~
4\alpha_{\ol z}=i a^{-1}\del_{\ol z}(ac)\ .
\label{alcpx}
\end{eqnarray}

\section{Gamma matrices}
\label{Gamma-matrices}

In this paper, we have chosen the $SO(1,9)$ gamma matrices $\Gamma^I$ to
be real $32\times 32$ matrices.\footnote{
In this appendix, we assume that the metric is flat.}
They can be written as
\begin{eqnarray}
 \Gamma^0=i\sigma_2\otimes 1_{16}\ ,~~~
 \Gamma^{i}=\sigma_1\otimes \gamma_{SO(8)}^{i}\ ,~~
(i=1,2,\cdots,9)
\end{eqnarray}
where $\gamma_{SO(8)}^{1\sim 8}$ are $SO(8)$ gamma matrices and
$\gamma_{SO(8)}^{9}$ is the chirality operator.\footnote{See, \emph{e.g.},
Appendix 5.B of \cite{Green:1987sp} for an explicit realization.}
$\Gamma^0$ is an anti-symmetric matrix and $\Gamma^{1\sim 9}$
are symmetric matrices.
$\Gamma^0 \Gamma^{I_1I_2\cdots I_n}$
are symmetric for $n=1,2$ (mod 4) and
anti-symmetric for $n=0,3$ (mod 4).

The following formulas are useful:
\begin{eqnarray}
\Gamma^{I}\Gamma^J
&=&\Gamma^{IJ}+\eta^{IJ}\ ,\\
 \Gamma^{IJ}\Gamma^K
&=&\Gamma^{IJK}+\eta^{JK}\Gamma^{I}-\eta^{IK}\Gamma^{J}\ ,
\\
 \Gamma^{IJK}\Gamma^L
&=&\Gamma^{IJKL}+\eta^{KL}\Gamma^{IJ}
-\eta^{JL}\Gamma^{IK}+\eta^{IL}\Gamma^{JK}\ ,\\
\Gamma^{IJ}\Gamma_{KL}
&=&
\Gamma^{IJ}_{~~KL}
-4\delta^{[I}_{[K}\Gamma^{J]}_{~L]}
-2\delta^{[I}_{[K}\delta^{J]}_{L]}
\ ,
\\
\Gamma^{IJK}\Gamma_{LM}
&=&
\Gamma^{IJK}_{~~~~LM}
+6\delta^{[I}_{[L}\Gamma^{JK]}_{~~~M]}
-6\delta^{[I}_{[L}\delta^J_{M]}\Gamma^{K]}
\ , 
\\
\Gamma_{LM}\Gamma^{IJK}
&=&
\Gamma_{LM}^{~~~~IJK}
-6\delta^{[I}_{[L}\Gamma^{JK]}_{~~~M]}
-6\delta^{[I}_{[L}\delta^J_{M]}\Gamma^{K]}
\ ,
\\
\Gamma^{IJK}\Gamma_{LMN}&=&\Gamma^{IJK}_{~~~~LMN}
+9\,\delta_{[L}^{[I}\Gamma^{JK]}_{~~MN]}
-18\,\delta_{[L}^{[I}\delta_{M}^{J}\Gamma^{K]}_{~N]}
-6\,\delta^{[I}_{L}\delta^J_M\delta^{K]}_{N}\ .
\end{eqnarray}
\begin{eqnarray}
 \Gamma^{I_1I_2\cdots I_n}\Gamma_{J_1J_2\cdots J_m}
&=&\Gamma^{I_1I_2\cdots I_n}_{~~~~~~~~J_1J_2\cdots J_m}
+(-1)^{n-1}nm\,\delta^{[I_1}_{[J_1}
\Gamma^{I_2\cdots I_n]}_{~~~~~~J_2\cdots J_m]}
\nn\\
&&+(-1)^{(n-1)+(n-2)}\frac{n(n-1)m(m-1)}{2}\,
\delta^{[I_1}_{[J_1}\delta^{I_2}_{J_2}
\Gamma^{I_3\cdots I_n]}_{~~~~~~J_3\cdots J_m]}+\cdots
\nn\\
&=&
\sum_{k=0}^{{\rm min}\{m,n\}}
(-1)^{(2n-k-1)k/2}k!{\,}_n\!C_k{\,}_m\!C_k
\,\delta^{[I_1}_{[J_1}\cdots
\delta^{I_k}_{J_k}
\Gamma^{I_{k+1}\cdots I_n]}_{~~~~~~~~J_{k+1}\cdots J_m]}
\ .
\end{eqnarray}

\begin{eqnarray}
\Gamma^{I'}\Gamma^{I_1\cdots I_n} \Gamma_{I'}
&=&(-1)^n(D-2n)\Gamma^{I_1\cdots I_n}\ ,\\
\Gamma^{I'J'}\Gamma^{I_1\cdots I_n} \Gamma_{J'I'}
&=&((D-2n)^2-D)\Gamma^{I_1\cdots I_n}\ .
\end{eqnarray}
where $D$ is the number of dimensions.
These formulas work for any $D$, though we are mainly interested in the
case $D=10$.

\section{Useful local transformations}

\subsection{Weyl transformation}
\label{Weyl}

Let us consider the transformations
\begin{eqnarray}
g_{\mu\nu}&\ra& e^{-2\omega} g_{\mu\nu}\ ,
\nn\\
e_\mu^{\hat\mu}&\ra&e^{-\omega} e_\mu^{\hat\mu}\ ,
\nn\\
\Gamma^\mu&\ra&e^{\omega}\Gamma^\mu\ ,
\nn\\
A_\mu&\ra& A_\mu\ ,
\nn\\
A_A&\ra&e^{\omega} A_A\ ,
\nn\\
\Psi&\ra&e^{\frac{3}{2}\omega}\Psi\ ,
\label{Weyl1}
\end{eqnarray}
where $\omega=\omega(x^\mu)$ is a real function of $x^\mu$.

Then the following quantities entering the action transform as:
\begin{eqnarray}
 D_\mu A^A&\ra& e^{\omega} (D_\mu A^A 
+ \del_\mu\omega\, A^A)\ ,
\\
 g^{\mu\mu'}\tr( D_\mu A^A D_{\mu'}A_A)
&\ra&
e^{4\omega}
g^{\mu\mu'}\tr( D_\mu A^A D_{\mu'}A_A
+2\del_\mu\omega\,(D_{\mu'}A^A) A_A
+\del_\mu\omega\del_{\mu'}\omega\,A^AA_A
)
\nn\\
&&=
e^{4\omega}
g^{\mu\mu'}\tr( D_\mu A^A D_{\mu'}A_A
+\del_\mu\omega\,D_{\mu'}(A^A A_A)
+\del_\mu\omega\del_{\mu'}\omega\, A^AA_A
)\ .
\nn\\
\end{eqnarray}
According to the definition
\begin{eqnarray}
D_{\mu}\Psi=\del_\mu\Psi+i[A_\mu,\Psi]
+\frac{1}{4}\omega_{\mu\hat\nu\hat\rho}\Gamma^{\hat\nu\hat\rho}\Psi\ ,
\end{eqnarray}
where the spin connection is 
\begin{eqnarray}
 \omega_{\mu\hat\nu\hat\rho}=\half e^{\nu'}_{\hat\nu}(\del_\mu e_{\nu'\hat\rho}
-(\del_{\nu'} g_{\mu\mu'})e^{\mu'}_{\hat\rho})-(\hat\nu\lra\hat\rho)\ ,
\end{eqnarray}
these objects transform as follows:
\begin{eqnarray}
\omega_{\mu\hat\nu\hat\rho}&\ra& \omega_{\mu\hat\nu\hat\rho}
+(e_{\mu\hat\rho}e^{\nu'}_{\hat\nu}
-e_{\mu\hat\nu}e^{\nu'}_{\hat\rho})\del_{\nu'}\omega\ ,
\\
D_{\mu}\Psi
&\ra&
e^{\frac{3}{2}\omega}\left(D_{\mu}
+\frac{3}{2}\del_\mu\omega
+\frac{1}{4}\del_{\nu'}\omega
(e_{\mu\hat\rho}e^{\nu'}_{\hat\nu}
-e_{\mu\hat\nu}e^{\nu'}_{\hat\rho})
\Gamma^{\hat\nu\hat\rho}
\right)\Psi
\nn\\
&&=e^{\frac{3}{2}\omega}\left(D_{\mu}
+\frac{1}{2}\del_{\nu'}\omega\,
(\Gamma^{\nu'}_{~\mu}+3\,\delta^{\nu'}_{\mu})
\right)\Psi\ .
\end{eqnarray}
Therefore
\begin{eqnarray}
 \Gamma^\mu D_{\mu}\Psi\ra
e^{\frac{5}{2}\omega}\, \Gamma^\mu D_{\mu}\Psi\ ,
\end{eqnarray}
and
\begin{eqnarray}
 \ol\Psi\Gamma^\mu D_{\mu}\Psi
&\ra&e^{4\omega}\, \ol\Psi\Gamma^\mu D_{\mu}\Psi\ .
\end{eqnarray}

The action (\ref{action1}) is invariant under the Weyl transformation
(\ref{Weyl1}), if we also transform the couplings as
\begin{eqnarray}
 d^{\mu\nu A}&\ra& e^{3\omega} d^{\mu\nu A}\ , 
\nn\\
 d^{\mu BA}&\ra& e^{2\omega} d^{\mu BA}\ , 
\nn\\
 d^{BCA}&\ra& e^{\omega} d^{BCA}\ , 
\nn\\
m^{AB}&\ra&e^{2\omega}m^{AB}+2e^{-2\omega}
g^{\mu\mu'}(-\del_\mu\omega\del_{\mu'}\omega
+a^{-1}D_\mu(a\del_{\mu'}\omega)) \delta^{AB}\ , 
\nn\\
M&\ra&e^{\omega} M\ .
\label{Weyl2}
\end{eqnarray}

\subsection{$SO(6)_R$ transformation}
\label{SO6tr}

Let us consider a local $SO(6)_R$ transformation
\begin{eqnarray}
A_A &\ra& O_A^{~\,B}A_B\ ,
\\
\Psi&\ra&\cO\Psi\ ,
\end{eqnarray}
where $O=(O^{~\,B}_A)\in SO(6)$ and $\cO$ is the corresponding $SO(6)$
element in the spinor representation of $SO(1,9)$ acting on the fermion.
This implies
\begin{eqnarray}
 D_\mu A_A&\ra&
 O_A^{~\,B}(D_\mu A_C+(O^{-1}\del_\mu O)_B^{~\,C}A_C)
\ ,
\\
g^{\mu\mu'}\tr( D_\mu A^AD_{\mu'} A_A)
&\ra&
g^{\mu\mu'}\tr( D_\mu A^AD_{\mu'}A_A
+2(O^{-1}\del_{\mu'} O)^{AC}(D_\mu A_A)A_C
\nn\\
&&
-(O^{-1}\del_{\mu'} OO^{-1}\del_{\mu'} O)^{AB}A_AA_B)
\ ,
\\
D_\mu\Psi&\ra&
\cO(D_\mu\Psi+\cO^{-1}\del_\mu \cO\Psi)
\ .
\end{eqnarray}

Then, the action (\ref{action1}) is invariant if we also
transform the couplings as
\begin{eqnarray}
d^{\mu\nu A}&\ra& d^{\mu\nu B}(O^{-1})_B^{~\,A}\ ,
\\
d_{\mu B}^{~~~A}&\ra&
 O_B^{~\,B'}d_{\mu B'}^{~~~A'}(O^{-1})_{A'}^{~\,A}+(O\del_\mu
 O^{-1})_{B}^{~\,A}\ ,
\\
d^{ABC}&\ra&d^{A'B'C'}(O^{-1})_{A'}^{~\,A}(O^{-1})_{B'}^{~\,B}(O^{-1})_{C'}^{~\,C}
\ ,
\\
m_A^{~\,B}&\ra&O_A^{~\,A'} m_{A'}^{~\,B'}(O^{-1})_{B'}^{~\,B} 
-2g^{\mu\nu}(O\del_\mu O^{-1}O\del_\nu O^{-1})_A^{~\,B}
\nn\\
&&
-2g^{\mu\nu}
(O_{A}^{~\,A'}d_{\mu A'}^{~~~B'}\del_\nu (O^{-1})_{B'}^{~~B}
-\del_\mu O_{A}^{~\,A'}d_{\nu A'}^{~~~B'}(O^{-1})_{B'}^{~~B})\ ,
\\
M&\ra&\cO M\cO^{-1}-\Gamma^\mu\cO\del_\mu\cO^{-1}\ .
\end{eqnarray}

Since $d_{\mu A}^{~~~B}$ behaves as the gauge field of the $SO(6)$
symmetry, it is useful to define the covariant derivative including
the $SO(6)$ gauge field:
\begin{eqnarray}
 \wh D_\mu A_A&=&D_\mu A_A+d_{\mu A}^{~~~B}A_B\ ,
\\
\wh D_\mu \Psi&=&
 D_\mu \Psi+\frac{1}{4}d_{\mu}^{~AB}\Gamma_{AB}\ ,
\label{so6spinor}
\end{eqnarray}
which transform as
\begin{eqnarray}
\wh D_\mu  A_A &\ra&O_A^{~\,B}\wh D_\mu A_B\ ,
\\
\wh D_\mu \Psi&\ra&\cO \wh D_\mu\Psi\ .
\end{eqnarray}
The action (\ref{action1}) can be written as
\begin{multline}
S =\int d^4 x \sqrt{-g_4}\,a
\tr
\bigg\{
-\frac{1}{2}
F_{\mu\nu}F^{\mu\nu}
-(\wh D_{\mu} A_{A})^2
+\half [A_{A},A_{B}]^2
+i(\ol\Psi\Gamma^\mu\wh D_\mu\Psi
+\ol\Psi\Gamma^{A}\,i [A_{A},\Psi])
\\
+\frac{c}{4}\,
\epsilon^{\mu\nu\rho\sigma}F_{\mu\nu}F_{\rho\sigma}
-d^{\mu\nu A}F_{\mu\nu}A_{A}
-d^{BCA}i[A_{B},A_{C}]A_{A}
-\frac{\wh m^{AB}}{2}
A_{A}A_{B}
-i\ol\Psi\wh M\Psi
\bigg\}\ ,
\label{action6}
\end{multline}
where
\begin{eqnarray}
\wh m^{AB}
&=& m^{AB}+2g^{\mu\nu}d_{\mu}^{~AA'}d_{\nu A'}^{~~~B}
\ ,\\
\wh M
&=&M+\frac{1}{4}\Gamma^{\mu AB}d_{\mu AB}\ .
\end{eqnarray}
$\wh m^{AB}$ and $\wh M$ transform as
\begin{eqnarray}
 \wh m_A^{~\,B}&\ra& (O\,\wh m\,O^{-1})_A^{~\,B}\ ,
\\
\wh M&\ra& \cO\wh M\cO^{-1}\ .
\end{eqnarray}

\section{Derivation of the SUSY conditions}
\label{appSUSYcond}

\subsection{Derivation of (\ref{SUSYcond1}) and (\ref{SUSYcond2})}
\label{derivation1}

First, the variation of the action (\ref{action0}) is given by
\begin{align}
\delta S
=&\
\int d^4 x \sqrt{-g}\,a
\tr
\Bigg\{
\delta A_{I}
\Big[
-2g^{II'}g^{JJ'}
D_J F_{I'J'}
+\left(-2a^{-1}D_\mu (a\,d^{I\mu A})
-m^{IA}\right) A_A
\nn\\
&\
\qquad\qquad\qquad\qquad\qquad
+\left(-2g^{IJ}g^{\mu K}a^{-1}\del_\mu a
-a^{-1}\del_\nu(ac)\,\epsilon^{\nu IJK}
-3d^{[IJK]}
\right)
F_{JK}
\Big]
\nn
\\
&\
\qquad\qquad\qquad\qquad
+
2i\,\delta\ol\Psi
\left(
\Gamma^I D_I\Psi
-\wt M
\Psi
\right)
-\ol\Psi\Gamma^I[\delta A_I,\Psi]
\Bigg\}
+(\mbox{total derivative terms})\ ,
\label{varS1}
\end{align}
with the understanding that
$d^{JKI}$ and $m^{IA}$ can be non-zero only if
$I=4\sim 9$, and $\epsilon^{IJKL}$ can be non-zero only for
$I,J,K,L=0\sim 3$.
Here, we have used
\begin{eqnarray}
\delta F_{IJ}=D_I\delta A_J-D_J\delta A_I\ ,~~~
\epsilon^{\mu\nu\rho\sigma}D_\nu F_{\rho\sigma}=0\ ,~~~
\ol\Psi\Gamma^I\delta\Psi=-\delta\ol\Psi\Gamma^I\Psi\ ,
\end{eqnarray}
and defined
\begin{eqnarray}
 d^{[IJK]}&\equiv& \frac{1}{3}(d^{IJK}+d^{JKI}+d^{KIJ})\ ,
\\
\wt M&\equiv& M-\frac{1}{2} \Gamma^\mu\del_\mu\log a\ .
\end{eqnarray}

The SUSY variation of the action with respect to the transformations
(\ref{SUSYtr}) is given by
\begin{align}
\delta_\epsilon S
=&\
\int d^4 x \sqrt{-g}\,a\,
\tr
\Bigg\{
(i\ol\epsilon\,\Gamma_I\Psi)
\Big[
-2g^{II'}g^{JJ'}
D_J F_{I'J'}
+\left(
-2a^{-1}D_\mu(a\,d^{I\mu A})
-m^{IA}\right) A_A
\nn
\\
&\
\qquad\qquad\qquad\qquad\qquad\qquad
+\left(-2g^{IJ}g^{\mu K}
a^{-1}\del_\mu a
-a^{-1}\del_\nu(ac)\,\epsilon^{\nu IJK}
-3d^{[IJK]}\right)
F_{JK}
\Big]
\nn\\
&\
\qquad\qquad\qquad\qquad
+
i\,\ol\epsilon\,(-F_{JK}\Gamma^{JK}+A_A\ol B^A)
\left(
\Gamma^I D_I\Psi
-\wt M \Psi
\right)
\Bigg\}
+(\mbox{total deriv. terms})\ ,
\label{susyS}
\end{align}
where we have used the identity
\footnote{See appendix 4.A of \cite{Green:1987sp} for a proof of this
identity.}
\begin{eqnarray}
\tr( \ol\Psi\Gamma^I[\delta_\epsilon A_I,\Psi])=0\ .
\end{eqnarray}

We would like to find a condition under which the integrand of
 (\ref{susyS}) is a total derivative.
Our ansatz for the total derivative is
\begin{eqnarray}
S_{\rm der}&=&
\int d^4x\sqrt{-g}\tr\left\{
D_I(i\ol\epsilon A^{IJK}\Psi F_{JK})
+D_I(i\ol\epsilon B^{IA}\Psi A_A)
\right\}\ ,
\end{eqnarray}
where $A^{IJK}$ and $B^{IA}$ are
$32\times 32$ matrices with
$A^{IJK}=-A^{IKJ}$ to be determined.
This total derivative term can be expanded as
\begin{eqnarray}
S_{\rm der} &=&
\int d^4x\sqrt{-g}\tr\bigg\{
i\ol\epsilon A^{IJK}(\Psi (D_I F_{JK})
+(D_I\Psi) F_{JK})+i D_I(\ol\epsilon B^{IA})\Psi A_A
\nn\\
&&
+i
\left(\ol\epsilon B^{[JK]}
+D_I(\ol\epsilon A^{IJK})\right)
\Psi F_{JK}
+i\ol\epsilon B^{IA}(D_I\Psi) A_A
\bigg\}\ .
\label{totalder2}
\end{eqnarray}

Comparing (\ref{susyS}) and (\ref{totalder2}),
we obtain the following conditions:
\begin{eqnarray}
\ol\epsilon \Gamma_I(-g^{II'}g^{JJ'}+g^{IJ'}g^{JI'})a
=\ol\epsilon (A^{JI'J'}+E^{JI'J'})
\ ,
\label{cond1}
\end{eqnarray}
\begin{eqnarray}
a\ol\epsilon(-\Gamma^{JK}\Gamma^I)=\ol\epsilon A^{IJK}
\ ,
\label{cond2}
\end{eqnarray}
\begin{eqnarray}
&&\ol\epsilon\Gamma_I\left(
(-g^{IJ}g^{\mu K}+g^{IK}g^{\mu J})\del_\mu a
+\del_\nu(ac)\,\epsilon^{I\nu JK}
-3a\,d^{[JKI]}\right)
+a\,\ol\epsilon \Gamma^{JK}\wt M
=
\ol\epsilon B^{[JK]} +D_I(\ol\epsilon A^{IJK})\ ,
\nn\\
\label{cond3}
\end{eqnarray}
\begin{eqnarray}
 a\ol\epsilon\ol{B}^A\Gamma^I=\ol\epsilon B^{IA}\ ,
\label{cond4}
\end{eqnarray}
\begin{eqnarray}
 \ol\epsilon\Gamma_I\left(-2 D_J(a\,d^{IJA})-a\,m^{IA}\right)
-a\ol\epsilon\ol{B}^A\wt M
=
D_I(\ol\epsilon B^{IA})\ .
\label{cond5}
\end{eqnarray}
In (\ref{cond1}), $E^{IJK}$ is a $32\times 32$ matrix
that is totally anti-symmetric with respect to
$I,J,K$, which can be added because of the Bianchi identity:
\begin{eqnarray}
 D_IF_{JK}+ D_JF_{KI}+ D_KF_{IJ}=0\ .
\end{eqnarray}

Equations (\ref{cond2}) and (\ref{cond4}) determine $\eb A^{IJK}$ and
 $\eb B^{IA}$, respectively. It is easy to see that (\ref{cond1})
is satisfied with $E^{IJK}=a\Gamma^{IJK}$, using the identity
\begin{eqnarray}
 \Gamma^{JK}\Gamma^I=\Gamma^{IJK}+g^{IK}\Gamma^J-g^{IJ}\Gamma^K\ .
\end{eqnarray}
Then, using (\ref{cond2}) and (\ref{cond4}),
 (\ref{cond3}) becomes
\begin{eqnarray}
&&
\ol\epsilon\left(\left(-\Gamma^Jg^{\mu K}+\Gamma^Kg^{\mu J}
\right)\del_\mu a
+\del_\nu(ac)\,\Gamma_I\epsilon^{I\nu JK}
-3a\,\Gamma_Id^{[JKI]}\right)
+a\,\ol\epsilon \Gamma^{JK}\wt M
\nn\\
&=&
a\ol\epsilon \ol{B}^{[K}\Gamma^{J]}
-D_I(
a\ol\epsilon(\Gamma^{JK}\Gamma^I)
)\ ,
\end{eqnarray}
which is equivalent to
\begin{eqnarray}
 (D_K\ol\epsilon)\Gamma^{IJ}\Gamma^K=
\ol\epsilon
\left(
 \ol{B}^{[J}\Gamma^{I]}
-
\Gamma^{IJ\mu}a^{-1}\del_\mu a
+(a^{-1}\del_\nu(ac)\,\epsilon^{\nu IJK}
+3\,d^{[IJK]})\Gamma_K
-\Gamma^{IJ}\wt M
\right)\ .
\end{eqnarray}
This is the condition (\ref{SUSYcond1}).
Similarly, (\ref{cond5}) becomes
\begin{eqnarray}
 \ol\epsilon\Gamma_I\left(-2 D_J(a\,d^{IJA})-a\,m^{IA}\right)
-a\ol\epsilon\ol{B}^A\wt M
=D_I(a\ol\epsilon\ol{B}^{A})\Gamma^I
\ ,
\end{eqnarray}
which is equivalent to
\begin{eqnarray}
a^{-1}D_I(a \ol\epsilon\ol{B}^{A})\Gamma^I
=\ol\epsilon
\left(-2a^{-1}\Gamma_ID_J(a\,d^{IJA})-m^{AB}\Gamma_B 
-\ol{B}^A\wt M
\right)
\ .
\end{eqnarray}
This gives (\ref{SUSYcond2}).

\subsection{Derivation of (\ref{SUSYcond1-e})$\sim$(\ref{Dmueq2})}
\label{derivation2}

Using the identity
\begin{eqnarray}
\Gamma^{IJ}\Gamma^K
\left(
\frac{1}{16}
\Gamma_{[I}g_{J]L}
-\frac{7}{16\times 54} 
\Gamma_{IJ}
\Gamma_{L}
\right)=\delta^K_L\ ,
\end{eqnarray}
we rewrite (\ref{SUSYcond1}) as
\begin{eqnarray}
D_L\ol\epsilon
=
\ol\epsilon\,C^{IJ}
\left(
\frac{1}{16}
\Gamma_{[I}g_{J]L}
-\frac{7}{16\times 54} 
\Gamma_{IJ}
\Gamma_{L}
\right)\ ,
\label{SUSYcond1-1-2}
\end{eqnarray}
where
\begin{eqnarray}
C^{IJ}
\equiv
 \ol{B}^{[J}\Gamma^{I]}
-\Gamma^{IJ\mu}a^{-1}\del_\mu a
+(a^{-1}\del_\nu(ac)\,\epsilon^{\nu IJK}
+3\,d^{[IJK]})\Gamma_{K}
-\Gamma^{IJ}\wt M
\ .
\end{eqnarray}

Inserting (\ref{SUSYcond1-1-2}) back into (\ref{SUSYcond1}), we obtain
\begin{eqnarray}
\ol\epsilon\, C^{IJ}
=
\ol\epsilon\,C^{I'J'}P_{I'J'}^{~~~IJ}
\ ,
\label{SUSYcond1-2-2}
\end{eqnarray}
where we have defined
\begin{eqnarray}
 P_{I'J'}^{~~~IJ}
&\equiv&
\left(
\frac{1}{16}
\Gamma_{[I'}g_{J']K}
-\frac{7}{16\times 54} 
\Gamma_{I'J'}
\Gamma_{K}
\right)
\Gamma^{IJ}\Gamma^K
\nn\\
&=&
\frac{1}{72}\Gamma_{I'J'}\Gamma^{IJ}
+\frac{1}{4}\Gamma_{[I'}\Gamma^{[I}\delta_{J']}^{J]}\ .
\label{proj-2}
\end{eqnarray}

One can check that $P_{I'J'}^{~~~IJ}$ is
 a projection operator satisfying
\begin{eqnarray}
 P_{IJ}^{~~I'J'} P_{I'J'}^{~~~KL}= P_{IJ}^{~~KL}\ ,
\end{eqnarray}
and
\begin{eqnarray}
 G^{[I'}\Gamma^{J']}P_{I'J'}^{~~~IJ}
=G^{[I}\Gamma^{J]}\ ,~~~
P_{I'J'}^{~~~IJ} G_{[I}\Gamma_{J]}
=G_{[I'}\Gamma_{J']}\ ,
\label{app-GGamP}
\end{eqnarray}
with arbitrary $G^J$.
Therefore, if $\ol\epsilon\, C^{IJ}$ can be written as
\begin{eqnarray}
\ol\epsilon\, C^{IJ} =\ol\epsilon\,G^{[I}\Gamma^{J]}\ ,
\label{CGGam}
\end{eqnarray}
with some $G^I$, (\ref{SUSYcond1-2-2}) is satisfied. Conversely,
if (\ref{SUSYcond1-2-2}) is satisfied,
\begin{eqnarray}
 G^I\equiv
  \frac{1}{72}(C^{I'J'}\Gamma_{I'J'})\Gamma^I+\frac{1}{4}C^{II'}\Gamma_{I'}
\end{eqnarray}
satisfies (\ref{CGGam}).
It can also be shown, using (\ref{app-GGamP}),
that $v^{I'J'}P_{I'J'}^{~~~IJ}=0$ is equivalent
to $v^{[IJ]}\Gamma_J=0$.

Then, using the property (\ref{app-GGamP}), one can easily show
\begin{eqnarray}
&&C^{I'J'}P_{I'J'}^{~~~IJ}- C^{IJ}
=
\left((a^{-1}\del_\nu(ac)\,\epsilon^{\nu I'J'K'}
+3\,d^{[I'J'K']})\Gamma_{K'}
-\Gamma^{I'J'}\wt M\right)
\left(
P_{I'J'}^{~~~IJ}
-\delta_{I'}^I\delta_{J'}^J
\right)
\ .
\nn\\
\end{eqnarray}
Therefore, (\ref{SUSYcond1-2-2}) can be written as
\begin{eqnarray}
0=\ol\epsilon \left((a^{-1}\del_\nu(ac)\,\epsilon^{\nu I'J'K'}
+3\,d^{[I'J'K']})\Gamma_{K'}
-\Gamma^{I'J'}\wt M\right)
\left(
P_{I'J'}^{~~~IJ}
-\delta_{I'}^I\delta_{J'}^J
\right)\ .
\label{SUSYcond1-2-3}
\end{eqnarray}
In this equation, $\wt M$ can be replaced with $M$,
because of the relation (\ref{app-GGamP}).

When $M$ is expanded as
\begin{eqnarray}
 M=m_{IJK}\Gamma^{IJK}\ ,
\end{eqnarray}
it satisfies
\begin{eqnarray}
 \Gamma^{I'J'}M
=M\Gamma^{I'J'}
-6 m_{IJK}
(g^{I'K}\Gamma^{IJ}\Gamma^{J'}
-g^{J'K}\Gamma^{IJ}\Gamma^{I'}
+4g^{I'I}g^{JJ'}\Gamma^{K})\ .
\end{eqnarray}
Therefore, (\ref{SUSYcond1-2-3}) is equivalent to
\begin{eqnarray}
0=\ol\epsilon\, e^{I'J'K'}
\Gamma_{K'}
\left(
P_{I'J'}^{~~~IJ}
-\delta_{I'}^I\delta_{J'}^J
\right)\ ,
\label{SUSYcond1-2-4}
\end{eqnarray}
where
\begin{eqnarray}
e^{IJK}\equiv a^{-1}\del_\nu(ac)\,\epsilon^{\nu IJK}
+3\,d^{[IJK]}+24\,m^{IJK}\ .
\end{eqnarray}
This is (\ref{SUSYcond1-e}).

Next, we write (\ref{SUSYcond1-1-2}) as
\begin{eqnarray}
 D_J\ol\epsilon=\ol\epsilon\,E_J\ ,
\label{DepE}
\end{eqnarray}
where
\begin{eqnarray}
E_J &\equiv&
\ol{B}^A\left(\frac{1}{4}\,g_{JA}
-\frac{1}{24}\Gamma_{A}\Gamma_J\right)
-\frac{1}{12}\Gamma^{\mu}\Gamma_J\del_\mu\log a
\nn\\
&&
+
e^{I'J'K'}
\left(\frac{1}{16}g_{JJ'}\Gamma_{K'I'}
-\frac{7}{16\times 54}\Gamma_{I'J'K'}\Gamma_J\right)
-m^{I'J'K'}
\left(3 g_{JJ'}\Gamma_{K'I'}
-\frac{1}{3}\Gamma_{I'J'K'}\Gamma_J\right)
\ .
\nn\\
\end{eqnarray}

(\ref{DepE}) with $J=A$, implies
\begin{eqnarray}
0&=&
\ol\epsilon\,
\Bigg(\ol{B}^{A'}\left(\frac{1}{4}\,\delta_{A'}^A
-\frac{1}{24}\Gamma_{A'}\Gamma^A\right)
-\frac{1}{12}\Gamma^{\mu A}
\del_\mu\log a
\nn\\
&&
+
e^{I'J'K'}
\left(\frac{1}{16}\delta^A_{J'}\Gamma_{K'I'}
-\frac{7}{16\times 54}\Gamma_{I'J'K'}\Gamma^A\right)
-m^{I'J'K'}
\left(3\delta^A_{J'}\Gamma_{K'I'}
-\frac{1}{3}\Gamma_{I'J'K'}\Gamma^A\right)
\Bigg)\ .
\nn\\
\label{B1}
\end{eqnarray}
This equation can be written as
\begin{eqnarray}
\ol\epsilon\,
\left(\ol{B}^{A}+\left(\frac{1}{4}e^{AK'I'}-12m^{AK'I'}
\right)\Gamma_{K'I'}
\right)
=\ol\epsilon\, F \Gamma^A\ ,
\label{Beq2-2}
\end{eqnarray}
where
\begin{eqnarray}
F\equiv
\frac{1}{6}\ol{B}^{A'}\Gamma_{A'}
+\frac{7}{4\times 54}e^{I'J'K'}\Gamma_{I'J'K'}
-\frac{4}{3}m^{I'J'K'}\Gamma_{I'J'K'}
+\frac{1}{3}\Gamma^{\mu}\del_\mu\log a\ . 
\label{F}
\end{eqnarray}
This is (\ref{Beq2}).
Multiplying (\ref{B1}) by $\Gamma_A$, we obtain
\begin{eqnarray}
 0=\ol\epsilon\,\bigg(\frac{1}{72}e^{IJK}\Gamma_{IJK}
-\frac{1}{2}\Gamma^{\mu}\del_\mu\log a
-\left(
\frac{1}{16}e_{\mu JK}-3m_{\mu JK}\right)\Gamma^{\mu JK}
-M
\bigg)\ ,
\label{am2-2}
\end{eqnarray}
which is (\ref{am2}).

With (\ref{am2-2}), the explicit form of $F$ in (\ref{F}) is not
needed anymore, because (\ref{am2-2}) implies that (\ref{F}) can
be replaced with
\begin{eqnarray}
F=
\frac{1}{6}
\left(\ol{B}^{A}+\left(\frac{1}{4}e^{AK'I'}-12m^{AK'I'}
\right)\Gamma_{K'I'}
\right)\Gamma_A
\ ,
\label{F2}
\end{eqnarray}
which can be obtained by contracting
(\ref{Beq2-2}) with  $\Gamma_A$. 

(\ref{DepE}) with $J=\mu$ is written as
\begin{eqnarray}
D_\mu\ol\epsilon
=
\ol\epsilon\,\left(
-\frac{1}{4}F\Gamma_\mu
+
\left(\frac{1}{16}e_{\mu IJ}
-3m_{\mu IJ}
\right)
\Gamma^{IJ}
\right)\ .
\end{eqnarray}
This equation is equivalent to (\ref{Dmueq2}).

\section{On $e^{IJK}$}
\label{app-eIJK}

\subsection{$e^{I\alpha a}=0$ $\Rightarrow$ $e^{IJK}=0$}
\label{e-mixed}

In this section, we split the 10-dimensional gamma matrices
$\{\Gamma^I\}$ into two sectors
$\{\Gamma^\alpha\}$ and $\{\Gamma^a\}$,
where $I,J,K=0,1,\cdots,9$;
 $\alpha,\beta,\gamma=0,1,\cdots, d-1$
and $a,b,c= d,d+1,\cdots, 9$, with $0<d<10$.
For simplicity, the 10 dimensional metric is assumed to be the flat
Minkowski metric, though the generalization to a curved metric is
straightforward.

Here, we prove the following statement: Suppose that $e^{IJK}$ with mixed
indices such as $e^{\alpha\beta a}$ and $e^{\alpha ab}$ are all zero.
Then, the condition
(\ref{SUSYcond1-e}) implies that all the components of $e^{IJK}$ vanish
when $d\ne 3,7$. This also holds for the case with $d=7$, which will be
shown separately in Appendix \ref{e-ISO3}.

The condition (\ref{SUSYcond1-e}) is equivalent to
\begin{eqnarray}
 0=\eb\,\left(e^{I'J'K'}\Gamma_{I'J'K'}\Gamma^{IJ}
+9e^{K'I'J}\Gamma_{K'I'}\Gamma^{I}
-9e^{K'I'I}\Gamma_{K'I'}\Gamma^{J}
-72e^{IJK'}\Gamma_{K'}\right)\ .
\end{eqnarray}
When $e^{\alpha\beta a}=0$ and $e^{\alpha a b}=0$,
this condition for $(I,J)=(\alpha,\beta)$,
$(I,J)=(a,b)$ and $(I,J)=(\alpha,a)$ becomes, respectively,
\begin{eqnarray}
0&=&\eb\left(e^{I'J'K'}\Gamma_{I'J'K'}\Gamma^{\alpha\beta}
+9e^{\gamma\delta\beta}\Gamma_{\gamma\delta}\Gamma^{\alpha}
-9e^{\beta'\gamma'\alpha}\Gamma_{\beta'\gamma'}\Gamma^{\beta}
-72e^{\alpha\beta\gamma'}\Gamma_{\gamma'}\right)\ ,
\label{e-1}
\\
0&=&\eb\left(e^{I'J'K'}\Gamma_{I'J'K'}\Gamma^{ab}
+9e^{b'c'b}\Gamma_{b'c'}\Gamma^{a}
-9e^{b'c'a}\Gamma_{b'c'}\Gamma^{b}
-72e^{abc'}\Gamma_{c'}\right)\ ,
\label{e-2}
\\
0&=&\eb\left(e^{I'J'K'}\Gamma_{I'J'K'}\Gamma^{\alpha a}
+9e^{b'c'a}\Gamma_{b'c'}\Gamma^{\alpha}
-9e^{\beta'\gamma'\alpha}\Gamma_{\beta'\gamma'}\Gamma^{a}
\right)\ ,
\label{e-3}
\end{eqnarray}
respectively.
Multiplying (\ref{e-1}) by $\Gamma_{\alpha\beta}$ we obtain
\begin{eqnarray}
0=\eb\,\left(d(d-1)e^{abc}\Gamma_{abc}
+(9-d)(10-d)e^{\alpha\beta\gamma}\Gamma_{\alpha\beta\gamma}
\right)\ ,
\label{e-1Gamalbe}
\end{eqnarray}
and we have
\begin{eqnarray}
\eb\, e^{IJK}\Gamma_{IJK}
=\frac{18(d-5)}{d(d-1)}\,
\eb\, e^{\alpha\beta\gamma}\Gamma_{\alpha\beta\gamma}
=\frac{18(5-d)}{(9-d)(10-d)}\,\eb\, e^{abc}\Gamma_{abc}\ .
\label{eG}
\end{eqnarray}
Inserting this equation into (\ref{e-1}) and (\ref{e-2}), we obtain
\begin{eqnarray}
0=\eb\left(\frac{2(5-d)}{(9-d)(10-d)}e^{a'b'c'}
\Gamma_{a'b'c'}\Gamma^{ab}
+e^{b'c'b}\Gamma_{b'c'}\Gamma^{a}
-e^{b'c'a}\Gamma_{b'c'}\Gamma^{b}
-8e^{abc'}\Gamma_{c'}\right)\ .
\label{e-2-2}
\end{eqnarray}

Multiplying  (\ref{e-2-2}) by $\Gamma_b$, we obtain
\begin{eqnarray}
 0=\eb\left(e^{a'b'c'}\G_{a'b'c'}\Gamma^a
-(10-d) e^{ab'c'}\Gamma_{b'c'}\right) \ ,
\label{e-1-2-1}
\end{eqnarray}
which implies
\begin{eqnarray}
0=\eb\left(2e^{a'b'c'}\G_{a'b'c'}
\Gamma^{ab}
+(10-d) e^{bb'c'}\Gamma_{b'c'}\Gamma^a
-(10-d) e^{ab'c'}\Gamma_{b'c'}\Gamma^b
\right) \ .
\end{eqnarray}
Inserting this back into (\ref{e-2-2}), we get
\begin{eqnarray}
0&=&\eb\left(
e^{bb'c'}\Gamma_{b'c'}\Gamma^{a}
-e^{ab'c'}\Gamma_{b'c'}\Gamma^{b}
-2(9-d)e^{abc'}\Gamma_{c'}\right)\ ,
\label{e-1-2-2}
\\
0&=&\eb\left(e^{a'b'c'}
\Gamma_{a'b'c'}\Gamma^{ab}
+(9-d)(10-d)e^{abc'}\Gamma_{c'}\right)\ .
\label{e-1-2-3}
\end{eqnarray}
{}From these equations, it is easy to show
\begin{eqnarray}
0&=&\eb\left(e^{abc} e_{ab'c'}
\Gamma_{bc}\Gamma^{b'c'}
+(9-d)e^{abc}e_{abc}
\right) \ ,
\label{e-1-3}
\\
0&=&\eb\left((e^{a'b'c'}
\Gamma_{a'b'c'})^2
+(9-d)(10-d)e^{abc}e_{abc}\right)\ .
\label{e-1-4}
\end{eqnarray}

Using the identities
\begin{eqnarray}
 \{\Gamma^{ab}, \Gamma_{a'b'}\}
&=&2\left(\Gamma^{ab}_{~~a'b'}-2
\delta^{[a}_{a'}\delta^{b]}_{b'}\right)\ ,
\label{Gam2Gam2}
\\
\{\Gamma^{abc},\Gamma_{a'b'c'}\}
&=&
2\left(9\,\Gamma^{[ab}_{~~[a'b'}\delta^{c]}_{c']} 
-6\,\delta^{[a}_{a'}\delta^{b}_{b'}\delta^{c]}_{c'}
\right)\ ,
\label{Gam3Gam3}
\end{eqnarray}
(\ref{e-1-3}) and (\ref{e-1-4}) can be written as
\begin{eqnarray}
0&=&\eb\left(e^{abc} e_{ab'c'}
\Gamma_{bc}^{~~b'c'}
+(7-d)e^{abc}e_{abc}
\right) \ ,
\\
0&=&\eb\left(9\,e^{abc} e_{ab'c'}
\Gamma_{bc}^{~~b'c'}
+(12-d)(7-d)e^{abc}e_{abc}
\right) \ .
\end{eqnarray}
These equations imply
\begin{eqnarray}
 0=(3-d)(7-d)\,\eb\, e^{abc}e_{abc}\ ,
\end{eqnarray}
and hence we have $e^{abc}=0$ for $d\ne 3,7$.

Then, (\ref{e-1})--(\ref{e-3}) with $e^{abc}=0$ imply
\begin{eqnarray}
 \eb\, e^{\alpha\beta\gamma}\Gamma_{\alpha\beta\gamma}=0\ ,~~~
 \eb\, e^{\alpha\beta\gamma}\Gamma_{\beta\gamma}=0\ ,~~~
 \eb\, e^{\alpha\beta\gamma}\Gamma_{\gamma}=0\ .
\end{eqnarray}
Multiplying the last equation by
 $e^{\alpha\beta\gamma'}\Gamma_{\gamma'}$ without summing over $\alpha$
and $\beta$, we obtain
\begin{eqnarray}
-(e^{\alpha\beta0})^2
+\sum_{\gamma=1}^{d-1}(e^{\alpha\beta\gamma})^2=0
\end{eqnarray}
for all $\alpha,\beta=0,1,\dots,d-1$ to have non-zero $\eb$.
Choosing $\alpha=0$, this equation implies
$e^{0\beta\gamma}=0$ and then again using this equation, we see that
all the components $e^{\alpha\beta\gamma}$ must vanish.

\subsection{Proof of $e^{IJK}=0$ for the $ISO(3)$ symmetric case}
\label{e-ISO3}

Here, we show that $e^{IJK}=0$ is the only solution
of (\ref{SUSYcond1-e}) with $\eb\ne 0$ for the cases with
$ISO(3)$ symmetry considered in Section \ref{ISO3}.
Because of the $ISO(3)$ symmetry, the allowed components of
$e^{IJK}$ are $e^{ABC}$, $e^{0AB}$ and $e^{123}$,
 where $A, B, C=4,\dots,9$. This is the case with $d=7$
considered in appendix \ref{e-mixed} by assigning the indices as
$\alpha,\beta,\gamma=0,4,\dots,9$ and $a,b,c=1,2,3$

The condition \eqref{SUSYcond1-e} with
 $(I,J)=(1,2)$ implies 
\begin{align}
0= \eb \left( \f{1}{72} e^{I'J'K'} \G_{I'J'K'} +\half e^{123}\G_{123}
\right)
\ .
\label{eIJK1-ISO3} 
\end{align}
Using this equation, 
the condition \eqref{SUSYcond1-e} with
$(I,J)=(0,1)$ implies
\beal{
0 =\eb  \left( \f{3}{4} e^{123} \G_{123} +\f{1}{8} e^{0AB} \G_{0AB}
\right)
\ .
\label{eIJK2-ISO3}
}
Multiplying the condition \eqref{SUSYcond1-e} with $(I,J)=(1,C)$
by $\G_{1C}$ and summing over $C=4,\dots,9$, we obtain
\beal{
0= \eb \left(3 e^{123}\G_{123} +\f{1}{8} e^{ABC} \G_{ABC}  \right)
\ .
}
Because $\Gamma_{ABC}\Gamma_{123}$ is an anti-symmetric matrix,
this equation implies that $e^{123}$ is an eigenvalue
of a real anti-symmetric matrix. This is possible only when
\begin{eqnarray}
 \ e^{123} =0\ ,~~~  \eb\,  e^{ABC} \G_{ABC} =0\ . 
\end{eqnarray}
These equations, together with (\ref{eIJK1-ISO3}) and
(\ref{eIJK2-ISO3}), imply
\begin{eqnarray}
\eb\,e^{IJK}\Gamma_{IJK}=0\ ,~~~
\eb\,e^{0AB}\Gamma_{AB}=0\ .
\end{eqnarray}
Using these results, the condition (\ref{SUSYcond1-e}) with $(I,J)=(0,C)$
becomes
\beal{
0= \eb \left( -\f{3}{4} e^{0BC} \G_{0B}+\f{1}{8} e^{ABC} \G_{AB}
 \right)\ .
 \label{eIJK3-ISO3}
}
Multiplying the condition (\ref{SUSYcond1-e}) with $(I,J)=(B,C)$
by $\G_B$ and summing over $B$, we obtain
\begin{eqnarray}
\eb\, e^{ABC}\Gamma_{AB}=0\ .
\label{eGamAB}
\end{eqnarray}
This equation and (\ref{eIJK3-ISO3}) imply
\begin{eqnarray}
\eb\, e^{0AB}\Gamma_{B}=0\ . 
\end{eqnarray}
Multiplying this equation by $e^{0AC}\G_C$, without summing over the
index $A$, we obtain
\beal{
0= \eb \, \bigg(
\sum_{B=4}^6 (e^{0AB})^2 \bigg)\ . 
} 
Therefore, we get $e^{0AB}=0$ for all $A,B=4,\dots,9$.
Using this result and (\ref{eGamAB}), the condition (\ref{SUSYcond1-e}) with
$(I,J)=(B,C)$ implies
\begin{eqnarray}
 \eb\, e^{ABC}\Gamma_A=0\ .
\end{eqnarray}
Again, multiplying this equation by $e^{A'BC}\Gamma_{A'}$, we obtain
$e^{ABC}=0$.
In conclusion, all the components of $e^{IJK}$ are zero for this case.


\begin{thebibliography}{10}

\bibitem{Vafa:1996xn}
C.~Vafa, ``{Evidence for F theory},''
  \href{http://dx.doi.org/10.1016/0550-3213(96)00172-1}{{\em Nucl. Phys.} {\bf
  B469} (1996)  403--418},
\href{http://arxiv.org/abs/hep-th/9602022}{{\tt arXiv:hep-th/9602022
  [hep-th]}}.

\bibitem{Bak:2003jk}
D.~Bak, M.~Gutperle, and S.~Hirano, ``{A Dilatonic deformation of AdS(5) and
  its field theory dual},''
  \href{http://dx.doi.org/10.1088/1126-6708/2003/05/072}{{\em JHEP} {\bf 05}
  (2003)  072},
\href{http://arxiv.org/abs/hep-th/0304129}{{\tt arXiv:hep-th/0304129
  [hep-th]}}.

\bibitem{Clark:2004sb}
A.~B. Clark, D.~Z. Freedman, A.~Karch, and M.~Schnabl, ``{Dual of the Janus
  solution: An interface conformal field theory},''
  \href{http://dx.doi.org/10.1103/PhysRevD.71.066003}{{\em Phys. Rev.} {\bf
  D71} (2005)  066003},
\href{http://arxiv.org/abs/hep-th/0407073}{{\tt arXiv:hep-th/0407073
  [hep-th]}}.

\bibitem{Witten:1988ze}
E.~Witten, ``{Topological Quantum Field Theory},''
\href{http://dx.doi.org/10.1007/BF01223371}{{\em Commun. Math. Phys.} {\bf 117}
  (1988)  353}.

\bibitem{Festuccia:2011ws}
G.~Festuccia and N.~Seiberg, ``{Rigid Supersymmetric Theories in Curved
  Superspace},'' \href{http://dx.doi.org/10.1007/JHEP06(2011)114}{{\em JHEP}
  {\bf 06} (2011)  114},
\href{http://arxiv.org/abs/1105.0689}{{\tt arXiv:1105.0689 [hep-th]}}.

\bibitem{Clark:2005te}
A.~Clark and A.~Karch, ``{Super Janus},''
  \href{http://dx.doi.org/10.1088/1126-6708/2005/10/094}{{\em JHEP} {\bf 10}
  (2005)  094},
\href{http://arxiv.org/abs/hep-th/0506265}{{\tt arXiv:hep-th/0506265
  [hep-th]}}.

\bibitem{DHoker:2006vfr}
E.~D'Hoker, J.~Estes, and M.~Gutperle, ``{Ten-dimensional supersymmetric Janus
  solutions},'' \href{http://dx.doi.org/10.1016/j.nuclphysb.2006.08.017}{{\em
  Nucl. Phys.} {\bf B757} (2006)  79--116},
\href{http://arxiv.org/abs/hep-th/0603012}{{\tt arXiv:hep-th/0603012
  [hep-th]}}.

\bibitem{DHoker:2006qeo}
E.~D'Hoker, J.~Estes, and M.~Gutperle, ``{Interface Yang-Mills, supersymmetry,
  and Janus},'' \href{http://dx.doi.org/10.1016/j.nuclphysb.2006.07.001}{{\em
  Nucl. Phys.} {\bf B753} (2006)  16--41},
\href{http://arxiv.org/abs/hep-th/0603013}{{\tt arXiv:hep-th/0603013
  [hep-th]}}.

\bibitem{Gomis:2006cu}
J.~Gomis and C.~Romelsberger, ``{Bubbling Defect CFT's},''
  \href{http://dx.doi.org/10.1088/1126-6708/2006/08/050}{{\em JHEP} {\bf 08}
  (2006)  050},
\href{http://arxiv.org/abs/hep-th/0604155}{{\tt arXiv:hep-th/0604155
  [hep-th]}}.

\bibitem{DHoker:2007zhm}
E.~D'Hoker, J.~Estes, and M.~Gutperle, ``{Exact half-BPS Type IIB interface
  solutions. I. Local solution and supersymmetric Janus},''
  \href{http://dx.doi.org/10.1088/1126-6708/2007/06/021}{{\em JHEP} {\bf 06}
  (2007)  021},
\href{http://arxiv.org/abs/0705.0022}{{\tt arXiv:0705.0022 [hep-th]}}.

\bibitem{DHoker:2007hhe}
E.~D'Hoker, J.~Estes, and M.~Gutperle, ``{Exact half-BPS Type IIB interface
  solutions. II. Flux solutions and multi-Janus},''
  \href{http://dx.doi.org/10.1088/1126-6708/2007/06/022}{{\em JHEP} {\bf 06}
  (2007)  022},
\href{http://arxiv.org/abs/0705.0024}{{\tt arXiv:0705.0024 [hep-th]}}.

\bibitem{Gaiotto:2008sd}
D.~Gaiotto and E.~Witten, ``{Janus Configurations, Chern-Simons Couplings, And
  The theta-Angle in N=4 Super Yang-Mills Theory},''
  \href{http://dx.doi.org/10.1007/JHEP06(2010)097}{{\em JHEP} {\bf 06} (2010)
  097},
\href{http://arxiv.org/abs/0804.2907}{{\tt arXiv:0804.2907 [hep-th]}}.

\bibitem{Suh:2011xc}
M.~Suh, ``{Supersymmetric Janus solutions in five and ten dimensions},''
  \href{http://dx.doi.org/10.1007/JHEP09(2011)064}{{\em JHEP} {\bf 09} (2011)
  064},
\href{http://arxiv.org/abs/1107.2796}{{\tt arXiv:1107.2796 [hep-th]}}.

\bibitem{Kim:2008dj}
C.~Kim, E.~Koh, and K.-M. Lee, ``{Janus and Multifaced Supersymmetric
  Theories},'' \href{http://dx.doi.org/10.1088/1126-6708/2008/06/040}{{\em
  JHEP} {\bf 06} (2008)  040},
\href{http://arxiv.org/abs/0802.2143}{{\tt arXiv:0802.2143 [hep-th]}}.

\bibitem{Kim:2009wv}
C.~Kim, E.~Koh, and K.-M. Lee, ``{Janus and Multifaced Supersymmetric Theories
  II},'' \href{http://dx.doi.org/10.1103/PhysRevD.79.126013}{{\em Phys. Rev.}
  {\bf D79} (2009)  126013},
\href{http://arxiv.org/abs/0901.0506}{{\tt arXiv:0901.0506 [hep-th]}}.

\bibitem{Ganor:2010md}
O.~J. Ganor, Y.~P. Hong, and H.~S. Tan, ``{Ground States of S-duality Twisted
  N=4 Super Yang-Mills Theory},''
  \href{http://dx.doi.org/10.1007/JHEP03(2011)099}{{\em JHEP} {\bf 03} (2011)
  099},
\href{http://arxiv.org/abs/1007.3749}{{\tt arXiv:1007.3749 [hep-th]}}.

\bibitem{Ganor:2014pha}
O.~J. Ganor, N.~P. Moore, H.-Y. Sun, and N.~R. Torres-Chicon, ``{Janus
  configurations with SL(2,$\mathbb{Z}$)-duality twists, strings on mapping
  tori and a tridiagonal determinant formula},''
  \href{http://dx.doi.org/10.1007/JHEP07(2014)010}{{\em JHEP} {\bf 07} (2014)
  010},
\href{http://arxiv.org/abs/1403.2365}{{\tt arXiv:1403.2365 [hep-th]}}.

\bibitem{Martucci:2014ema}
L.~Martucci, ``{Topological duality twist and brane instantons in F-theory},''
  \href{http://dx.doi.org/10.1007/JHEP06(2014)180}{{\em JHEP} {\bf 06} (2014)
  180},
\href{http://arxiv.org/abs/1403.2530}{{\tt arXiv:1403.2530 [hep-th]}}.

\bibitem{Maxfield:2016lok}
T.~Maxfield, ``{Supergravity Backgrounds for Four-Dimensional Maximally
  Supersymmetric Yang-Mills},''
  \href{http://dx.doi.org/10.1007/JHEP02(2017)065}{{\em JHEP} {\bf 02} (2017)
  065},
\href{http://arxiv.org/abs/1609.05905}{{\tt arXiv:1609.05905 [hep-th]}}.

\bibitem{Lawrie:2016axq}
C.~Lawrie, S.~Sch\"afer-Nameki, and T.~Weigand, ``{Chiral 2d theories from N = 4
  SYM with varying coupling},''
  \href{http://dx.doi.org/10.1007/JHEP04(2017)111}{{\em JHEP} {\bf 04} (2017)
  111},
\href{http://arxiv.org/abs/1612.05640}{{\tt arXiv:1612.05640 [hep-th]}}.

\bibitem{Couzens:2017way}
C.~Couzens, C.~Lawrie, D.~Martelli, S.~Sch\"afer-Nameki, and J.-M. Wong,
  ``{F-theory and AdS$_{3}$/CFT$_{2}$},''
  \href{http://dx.doi.org/10.1007/JHEP08(2017)043}{{\em JHEP} {\bf 08} (2017)
  043},
\href{http://arxiv.org/abs/1705.04679}{{\tt arXiv:1705.04679 [hep-th]}}.

\bibitem{Assel:2016wcr}
B.~Assel and S.~Sch\"afer-Nameki, ``{Six-dimensional origin of $ \mathcal{N} =
  4$ SYM with duality defects},''
  \href{http://dx.doi.org/10.1007/JHEP12(2016)058}{{\em JHEP} {\bf 12} (2016)
  058},
\href{http://arxiv.org/abs/1610.03663}{{\tt arXiv:1610.03663 [hep-th]}}.

\bibitem{Harvey:2007ab}
J.~A. Harvey and A.~B. Royston, ``{Localized modes at a D-brane-O-plane
  intersection and heterotic Alice strings},''
  \href{http://dx.doi.org/10.1088/1126-6708/2008/04/018}{{\em JHEP} {\bf 04}
  (2008)  018},
\href{http://arxiv.org/abs/0709.1482}{{\tt arXiv:0709.1482 [hep-th]}}.

\bibitem{Buchbinder:2007ar}
E.~I. Buchbinder, J.~Gomis, and F.~Passerini, ``{Holographic gauge theories in
  background fields and surface operators},''
  \href{http://dx.doi.org/10.1088/1126-6708/2007/12/101}{{\em JHEP} {\bf 12}
  (2007)  101},
\href{http://arxiv.org/abs/0710.5170}{{\tt arXiv:0710.5170 [hep-th]}}.

\bibitem{Harvey:2008zz}
J.~A. Harvey and A.~B. Royston, ``{Gauge/Gravity duality with a chiral N=(0,8)
  string defect},'' \href{http://dx.doi.org/10.1088/1126-6708/2008/08/006}{{\em
  JHEP} {\bf 08} (2008)  006},
\href{http://arxiv.org/abs/0804.2854}{{\tt arXiv:0804.2854 [hep-th]}}.

\bibitem{Chu:2006pa}
C.-S. Chu and P.-M. Ho, ``{Time-dependent AdS/CFT duality and null
  singularity},'' \href{http://dx.doi.org/10.1088/1126-6708/2006/04/013}{{\em
  JHEP} {\bf 04} (2006)  013},
\href{http://arxiv.org/abs/hep-th/0602054}{{\tt arXiv:hep-th/0602054
  [hep-th]}}.

\bibitem{Das:2006dz}
S.~R. Das, J.~Michelson, K.~Narayan, and S.~P. Trivedi, ``{Time dependent
  cosmologies and their duals},''
  \href{http://dx.doi.org/10.1103/PhysRevD.74.026002}{{\em Phys. Rev.} {\bf
  D74} (2006)  026002},
\href{http://arxiv.org/abs/hep-th/0602107}{{\tt arXiv:hep-th/0602107
  [hep-th]}}.

\bibitem{Lin:2006ie}
F.-L. Lin and W.-Y. Wen, ``{Supersymmetric null-like holographic
  cosmologies},'' \href{http://dx.doi.org/10.1088/1126-6708/2006/05/013}{{\em
  JHEP} {\bf 05} (2006)  013},
\href{http://arxiv.org/abs/hep-th/0602124}{{\tt arXiv:hep-th/0602124
  [hep-th]}}.

\bibitem{Das:2006pw}
S.~R. Das, J.~Michelson, K.~Narayan, and S.~P. Trivedi, ``{Cosmologies with
  Null Singularities and their Gauge Theory Duals},''
  \href{http://dx.doi.org/10.1103/PhysRevD.75.026002}{{\em Phys. Rev.} {\bf
  D75} (2007)  026002},
\href{http://arxiv.org/abs/hep-th/0610053}{{\tt arXiv:hep-th/0610053
  [hep-th]}}.

\bibitem{Chu:2007um}
C.-S. Chu and P.-M. Ho, ``{Time-dependent AdS/CFT duality. II. Holographic
  reconstruction of bulk metric and possible resolution of singularity},''
  \href{http://dx.doi.org/10.1088/1126-6708/2008/02/058}{{\em JHEP} {\bf 02}
  (2008)  058},
\href{http://arxiv.org/abs/0710.2640}{{\tt arXiv:0710.2640 [hep-th]}}.

\bibitem{Awad:2007fj}
A.~Awad, S.~R. Das, K.~Narayan, and S.~P. Trivedi, ``{Gauge theory duals of
  cosmological backgrounds and their energy momentum tensors},''
  \href{http://dx.doi.org/10.1103/PhysRevD.77.046008}{{\em Phys. Rev.} {\bf
  D77} (2008)  046008},
\href{http://arxiv.org/abs/0711.2994}{{\tt arXiv:0711.2994 [hep-th]}}.

\bibitem{Awad:2008jf}
A.~Awad, S.~R. Das, S.~Nampuri, K.~Narayan, and S.~P. Trivedi, ``{Gauge
  Theories with Time Dependent Couplings and their Cosmological Duals},''
  \href{http://dx.doi.org/10.1103/PhysRevD.79.046004}{{\em Phys. Rev.} {\bf
  D79} (2009)  046004},
\href{http://arxiv.org/abs/0807.1517}{{\tt arXiv:0807.1517 [hep-th]}}.

\bibitem{Greene:1989ya}
B.~R. Greene, A.~D. Shapere, C.~Vafa, and S.-T. Yau, ``{Stringy Cosmic Strings
  and Noncompact Calabi-Yau Manifolds},''
\href{http://dx.doi.org/10.1016/0550-3213(90)90248-C}{{\em Nucl. Phys.} {\bf
  B337} (1990)  1--36}.

\bibitem{preparation}
J.~Choi, J.~J. Fernandez-Melgarejo, and S.~Sugimoto, ``{Deformation of ${\cal N}=4$ SYM with varying couplings via fluxes and intersecting branes},''
  \href{https://arxiv.org/abs/1801.09394}{{\tt arXiv:1801.09394 [hep-th]}} .

\bibitem{Green:1987sp}
M.~B. Green, J.~H. Schwarz, and E.~Witten, {\em {Superstring Theory. Vol. 1:
  Introduction}}.
\newblock Cambridge Monographs on Mathematical Physics.
1988.
\newblock

\end{thebibliography}

\providecommand{\href}[2]{#2}\begingroup\raggedright\endgroup

\end{document}